\documentclass[]{aa}   
\usepackage{footnote}
\usepackage{ulem}
\usepackage{graphicx}
\usepackage{enumerate}
\usepackage{natbib}
\usepackage[caption=false]{subfig}

\begin{document}

\title{Long-term study of Mkn 421 with the HAGAR Array of Telescopes}
\titlerunning{Long-term spectral and temporal study of Mkn 421}

\author{A. Sinha$^{1}$, A. Shukla$^{1,4}$, L. Saha$^{1,5}$, B. S. Acharya$^{1}$, G. C. Anupama$^{4}$, P. Bhattacharya$^{5}$, R. J. Britto$^{1,6}$, V. R. Chitnis$^{1}$, T. P. Prabhu$^{4}$, B. B. Singh$^{1}$ and P. R. Vishwanath$^{4}$}
\authorrunning{Sinha et. al.}

\institute{
\inst{1}~Tata Institute of Fundamental Research, Homi Bhabha Road, Colaba, Mumbai, 400 005, India.\\ 
\inst{2}~Indian Institute of Astrophysics, II Block, Koramangala, Bangalore, 560 034, India.\\
\inst{3}~Saha Institute of Nuclear Physics, 1/AF, Bidhannagar, Kolkata, 700 064, India.\\
\inst{4}~Now at ETH Zurich, Institute for Particle Physics, Otto-Stern-Weg 5, 8093 Zurich, Switzerland. \\
\inst{5}~Now at Nicolaus Copernicus Astronomical Center, Torun, 87-100, Poland.\\ 
\inst{6}~Now at Dept. of Physics, University of the Free State, PO Box 339, 9300 Bloemfontein, South Africa.\\
\email{atreyee@tifr.res.in} \\
}
\abstract
{ The HAGAR Telescope Array at Hanle, Ladakh has been regularly monitoring the nearby blazar Mkn 421 for the past 7yrs.}
{ Blazars show flux variability in all timescales across the electromagnetic spectrum. While there is abundant literature characterizing the short term flares from different blazars, comparatively little work has been done to study the long term variability. We aim to study the long term temporal and spectral variability in the radiation from Mkn 421 during 2009-2015.}
{We quantify the variability and lognormality from the radio to the VHE bands, and compute the correlations between the various wavebands using the z-transformed discrete correlation function. We construct the Spectral Energy Distribution (SED) contemporaneous with HAGAR observation seasons and fit it with a one zone synchrotron self Compton model to study the spectral variability.}
{ The flux is found to be highly variable across all time scales. The variability is energy dependant, and is maximum in the X-ray and Very High Energy (VHE) bands. A strong correlation is found between the Fermi-LAT (gamma) and radio bands, and between Fermi-LAT and optical, but none between Fermi-LAT and X-ray. Lognormality in the flux distribution is clearly detected. This is the third blazar, following BL~Lac and PKS~2155$+$304 to show this behavior. The SED can be well fit by a one zone SSC model, and variations in the flux states can be attributed mainly due to changes in the particle distribution. A strong correlation is seen between the break energy $\gamma_b$ of the particle spectrum and the total bolometric luminosity.}
{}
\keywords{BL Lacertae objects: individual (Mkn 421)- galaxies: active - X-rays: galaxies - radiation mechanisms: non-thermal}

\maketitle


\section[sect:Intro]{Introduction}

Blazars count among the most violent sources of high energy emission in the known universe. They are characterized by highly variable nonthermal emission across the entire electromagnetic spectrum, strong radio and optical polarization, and apparent superluminal motion. The likely explanation of such observations is that they are a subclass of active galactic nuclei (AGNs) where the emission originates from a relativistic jet oriented close to the line of sight \citep{UrryPadovani}. Blazars are the dominant sources of high energy radiation in the extragalactic sky, with over 50 blazars\footnote{www.tevcat.uchicago.edu} detected in the VHE regime. 
Blazars are further subdivided into BL~Lacertae (BL~Lac) and flat-spectrum
radio quasars (FSRQs), where BL~Lacs are characterized by the absence of (or very weak) emission lines. 
The broadband spectral energy distribution (SED) of blazars is characterized by two peaks, one in the IR - X-ray regime, and the second one in  the $\gamma$-ray regime. According to the location of the first peak, BL Lacs are further classified into low-energy peaked BL~Lacs (LBL), intermediate peaked BL~Lacs (IBL) and high-energy peaked BL~Lacs (HBL) \citep{PadGio}. Both leptonic and hadronic models have been used to explain the broadband SED with
varying degrees of success. The origin of the low-energy component is well established to be caused by synchrotron emission from
relativistic electrons gyrating in the magnetic field of the jet. However, the physical mechanisms responsible for the high-energy emission are
still under debate. It can be produced either through inverse Compton (IC) scattering of low-frequency photons by the same
electrons responsible for the synchrotron emission (leptonic models), or through hadronic processes initiated by relativistic
protons, neutral and charged pion decays, or muon cascades (hadronic models). The seed photons for IC in leptonic
models can be either the synchrotron photons themselves (synchrotron self-Compton, SSC) or from external sources such
as the broad line region (BLR), the accretion disc, and the cosmic microwave background (external Compton, EC). For a comprehensive review of
these mechanisms, see \cite{bottcher2007}. 

Mkn 421 is the closest ($z=0.031$) and the best-studied TeV blazar. It is also the first detected extragalactic TeV source \citep{punch}, and one of the brightest BL~Lac objects seen in the UV and X-ray bands. It is an HBL, with the synchrotron spectrum peaking in the X-ray regime. 
It is highly variable on all timescales and across all wavelengths from radio to TeVs, and has been well studied during its various flaring episodes by several authors \citet{MAGIC421_2008,MAGIC421_2006,MAGIC421_outburst,Amit421,MAXI421,Mkn421_2000,Suzaku421_2006,Swift421_2006,multi421_2005-2006,Integral421,421_march2001,MAGIC421,XMM421_2003, my421, vaidehi421,donna}. \citet{fermi421} studied the quiescent state emission from this source with the best-sampled SED to date. \citet{Liu_longterm_421} constructed the historical lightcurve of this object from 1900 to 1991 in the optical B-band, the analysis of which revealed a possible time period of $23.1 \pm 1.1$ yrs in the flux variations. \citet{VHE_longterm} compiled and studied the long term VHE light curve, and found significant correlation ($\sim 68\%$) with X-ray data from RXTE-ASM. Strong X-ray-TeV correlation has also been found at smaller time scales \citep{multi421,Katarzynski_hbl,Qian}. Recently, \cite{ybj421} have reported on extensive multiwavelength observations of this source, from 2008-2013. They found the VHE flux to be strongly correlated with the X-ray flux, but partially correlated with the GeV flux. Such moderate X-ray-GeV correlation is typical of most HBLs \citep{Xray-GeV_corr}.


The High Altitude GAmma Ray (HAGAR) telescope system at Hanle, India has been regularly monitoring Mkn 421 since its inception in October 2008. In this paper we present the results of the long term monitoring of this source for the past 7 years, from 2009-2015. We use data in the radio, optical, X-ray and gamma rays to characterize lognormality in the intrinsic flux variations across the multiwavelength spectrum. Correlation studies between different wavelengths are also performed. Spectral energy distribution during different flux states of the source have been constructed simultaneous with HAGAR observations, and fit with leptonic SSC models. The main aim of this paper is not to investigate the fast temporal and spectral variabilities, but to study the smooth variations over weekly and monthly timescales, which may throw light on the underlying jet environment.   

\section[sect:Analysis]{Multiwavelength observations and data analysis}\label{sect:Analysis}

During the period from 2009-2015, Mkn421 has been simultaneously observed by many instruments across the entire electromagnetic spectrum. Alongside data from HAGAR, we analyze data from the Fermi-LAT, {\it Nu}STAR,  {\it Swift} X-ray and UV telescopes. In addition, we include the X-ray observations by the Monitor of All-sky X-ray Image (MAXI) telescope, optical observations by SPOL CCD Imaging/Spectropolarimeter at Steward Observatory and radio data from the Owens Valley Radio Observatory (OVRO)  for the present study. The analysis procedures of these observations are described below. 

\subsection[sect:hagar]{HAGAR}\label{sect:HAGAR}

\begin{figure*}
	\centering
	\includegraphics[scale=0.4]{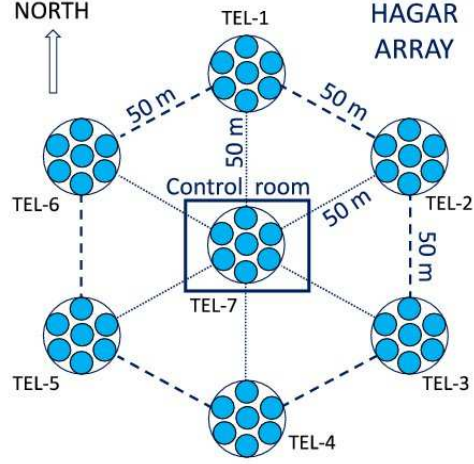}
	\caption{Schematic of the HAGAR telescope array. The hexagonal array has a radius of 50m, and each telescope has 7 PMTs in a hexagonal array.}
	\label{hag_schem}
\end{figure*}

The High Altitude GAmma Ray (HAGAR) array is a  hexagonal array of seven atmospheric Cherenkov telescopes which uses the wavefront sampling technique to detect celestial  gamma rays. It is located at the Indian Astronomical Observatory site (32$^\circ$ 46' 46" N, 78$^\circ$ 58' 35" E), in Hanle, Ladakh in the Himalayan mountain ranges at an altitude of 4270m. The Cherenkov photon density of a shower increases with altitude, and thus, a low energy  threshold is achieved by operating the telescope at such high altitudes. A schematic of the HAGAR telescope array is given in Fig. \ref{hag_schem}. The telescope structures are based on alt-azimuth design (details given in \cite{kiranHAGAR})  and the separation between two nearest telescopes is 50m.
Each of the seven telescopes has seven para-axially mounted front coated parabolic mirrors of diameter 0.9 m, with a UV-sensitive photo-multiplier tube (PMT) at the focus of individual mirrors. The opening angle (Field of View) of the HAGAR telescopes is 3$^{\circ}$. The pulses from the seven PMTs of each telescope are linearly added to form a telescope pulse for forming trigger and are brought to the control room through low attenuation co-axial cables. Data are recorded using a CAMAC based Data Acquisition system. The presence of at least four telescope pulses within a coincidence window of 60ns is used as the trigger for data recording and the relative arrival time of the Cherenkov shower front at each telescope is recorded for each event using a 12 bit Phillips Time to Digital Converter (TDC). A VME based DAQ  and and Acquiris digitizer has also been recently installed. Detials of the HAGAR array can be found in \cite{varshaicrc,Amit421}. 

Observations are carried out in ON-OFF mode, with a typical 60mins source run followed (or preceded) by a 60min background run with the same zenith angle. The data analysis is performed using a toolkit developed in-house using IDL based routines. The relative arrival time of the Cherenkov shower front at each telescope is used to reconstruct the arrival direction of the incident cosmic/$\gamma$-ray. The relative arrival times are corrected for fixed time offsets ($T0$) arising due to the difference in the
signal path length, propagation delay in processing electronics and transit time of
PMTs. $T0$s are calculated by pointing all the telescopes at a fixed angle (say 10$^{\circ}$ N) in a dark region of the sky, and minimizing the $\chi^2$ given by
\begin{equation}
	\label{equn:t0}
	\chi^2 = \Sigma w_{ij}(T0_i -T0_j - C_{ij})^2
\end{equation}

where $T0_i$ and $T0_j$ are the time offsets, $w_{ij}$ the weight factor, and $C_{ij}$ the mean delay between the $i^{th}$ and $j^{th}$ telescope. The arrival direction of the Cherenkov shower front is reconstructed by assuming it to be a plane wavefront. Various data quality cuts are imposed to get ``clean" data.

Detailed simulations for studying the performance parameters have been done in  \citet{labHAGAR}. The energy threshold for the HAGAR array for vertically incident $\gamma$-ray showers is 208 GeV, with a sensitivity of detecting a Crab nebula like source in 17 hr for a 5$\sigma$ significance. The integral flux of the Crab nebula above 208 GeV is $2.29 \times 10^{-6}$ ph/cm$^2$/sec, corresponding to a mean count rate of 6.6 counts/min. 

Mkn 421 has been regularly monitored by HAGAR since its setup in 2008. Analysis of all data till 2015 yields 101 hrs of ``clean" data. 
Monthly averaged $T0$s are used for reconstruction of arrival direction of the shower front, and the source counts computed according to the procedure outlined in \citet{Amit421}. Only events with signals in at least 5 telescopes are retained to minimize the systematic errors, for which the energy threshold is estimated to be $\sim 250 GeV$. Mkn 421 is detected at a mean level of $80\%$ of the Crab flux, at a significance of 9.7$\sigma$. The excess counts over background in each HAGAR season are listed in Table \ref{hagarcounts}. 

To study the spectral variations, multiwavelength SEDs are constructed contemporaneous with each HAGAR season.

\subsection[sect:fermi]{Fermi-LAT}\label{sect:fermi}

Fermi-LAT data from 2009-2015 are extracted from a region of 20$^{\circ}$ centered on the source. The standard data analysis procedure as mentioned in the Fermi-LAT  documentation\footnote{http://fermi.gsfc.nasa.gov/ssc/data/analysis/documentation/}
is used. Events belonging to the energy range 0.2$-$300 GeV and SOURCE class are used. To select good time intervals, 
a filter ``\texttt{DATA$\_$QUAL$>$0}'', \&\& ``\texttt{LAT$\_$CONFIG==1}'' is used and only events with less than 105$^{\circ}$ zenith angle are selected to avoid contamination from the Earth limb $\gamma$-rays. 
The galactic diffuse emission component gll\_iem\_v05\_rev1.fits and an isotropic component iso\_source\_v05\_rev1.txt are used as the background models. The unbinned likelihood method included in the pylikelihood library of {\tt Science Tools (v9r33p0)} and the post-launch instrument response functions P7REP\_SOURCE\_V15 are used for the analysis. All the sources lying within 10$^{\circ}$ region of interest (ROI) centered at the position of Mkn 421 and 
defined in the third {\it Fermi}-LAT catalog \citep{3FGL}, are included in the xml file. All the parameters except the scaling factor of the sources within the ROI are allowed to vary during the likelihood fitting. For sources between 10$^{\circ}$ to 20$^{\circ}$ from the centre, all parameters were kept frozen at the default values. The source is modelled by a simple power law. The light curve is binned over 10 days, and spectra are extracted in 6 logarithmically binned energy bins for the 21 different states contemporaneous with the HAGAR observation seasons.

\subsection{{\it Nu}STAR observations}\label{subsec:nustar}

   NuSTAR \citep{Nustar_instru} features the first focussing X-ray telescope to extend high sensitivity beyond 10 keV. NuStar observations for this source are available for the year 2013, amounting to a total of 24 pointings. The NuSTAR data are processed with the NuSTARDAS software package v.1.4.1 available within HEASOFT package (6.16). The latest CALDB (v.20140414) is used. 
After running {\tt nupipeline v.0.4.3} on each observation, {\tt nuproducts v.0.2.8} is used to obtain the light curves and spectra. 
Circular regions of 12 pixels centered on Mkn 421 and of 40 pixels centred on 165.96, 38.17  are used as source and background regions, respectively. The spectra from the two detectors A and B are combined using {\tt addascaspec}. Data from various observations are added using {\tt addascaspec} and then grouped (using the tool {\tt grppha v.3.0.1}) to ensure a minimum of 30 counts in each bin. Spectra are available for the states s10 - s14, and are analyzed simultaneously with spectra from Swift-XRT \citep{my421}.

\subsection{{\it Swift} observations}\label{subsec:swift}

There are 594 {\it Swift} pointings between January 2009 and June 2015. Publicly available daily binned source counts in the $15-50$ keV range are taken from the {\it Swift}-BAT webpage\footnote{http://swift.gsfc.nasa.gov/results/bs70mon/SWIFT\_J1104.4p3812}.

The XRT data \citep{XRT_instru} are processed with the XRTDAS software package (v.3.0.0) available within HEASOFT package (6.16). Event files are cleaned and calibrated using standard procedures ({\tt xrtpipeline v.0.13.0}), and {\tt xrtproducts v.0.4.2} is used to obtain the light curves and spectra.
 Standard grade selections of $0-12$ in the Windowed Timing (WT) mode are used. 
Circular regions of 20 pixels centred on Mkn 421 (at RA 166.113 and Dec 38.208) and of 40 pixels centred at 166.15, 38.17 are used as source and background regions, respectively.
For the observations affected by pileup (counts $> 100 cts/s$) \citep{XRT_pileup}, an annular region with an inner radius of 2 pixels and an outer radius of 20 pixels is taken as the source region. 
The light curves are finally corrected for telescope vignetting and PSF losses with the tool {\tt xrtlccorr v.0.3.8}. All pointings within each state are added using the tool {\tt addspec}. The spectra are grouped to ensure a minimum of 30 counts in each bin by using the tool {\tt grppha v.3.0.1}. 

Since the X-ray data lies at the peak of the synchrotron spectrum, significant departure from a simple power law is needed to model the  X-ray spectrum. Following \citep{my421}, we fit the observed spectrum with a log parabola given by
\begin{equation}
         dN/dE = K(E/E_b)^{-\alpha -\beta log(E/E_{b})}
 ,\end{equation}
\noindent
where $\alpha$ gives the spectral index at $E_{b}$. 
 The point of maximum curvature, $E_{p}$ is given by 
\begin{equation}
        \label{lp_phot}
        E_{p} = E_{b} 10^{(2-\alpha)/2\beta}
.\end{equation}
\noindent
During fitting, $E_{b}$ is fixed at $1keV$, and the  XRT and the NuSTAR (where available) spectral parameters are tied to each other, except for the relative normalization between the two instruments.  To correct for the line-of-sight absorption of soft X-rays due to the interstellar gas, the neutral hydrogen column density is fixed at $N_{H}= 1.92 \times 10^{20} cm^{-2} $ \citep{LAB}.

{\it Swift}-UVOT \citep{UVOT_instru} operated in imaging mode during this period, 
and for most of the observations, cycled through the UV filters UW1, UW2, and UM2. 
The tool {\tt uvotsource v.3.3} is used to extract the fluxes from each of the images using aperture photometry. 
The observed magnitudes are corrected for Galactic extinction ($E_{B-V}=0.019$ mag) using the dust maps of \cite{schelgel} and converted to flux units using the zero-point magnitudes and conversion factors of \citet{UVOT_conv}. 

\subsection{Other multiwavelength data}\label{subsec:other}
We supplement the above information with other available multiwavelength data, namely from:
\begin{enumerate}[(i)]
	\item {\bf MAXI} \\
		Publicly available daily binned source counts in the $2-20$ keV range from the  Monitor of All-sky X-ray Image (MAXI) on board the International Space Station (ISS) \citep{MAXI}  are taken from their website\footnote{http://maxi.riken.jp/}. MAXI orbital data are known to have some bad quality points due to fluctuations in the background or for a shadow of solar-battery paddles. However, this should not affect out results significantly since we use daily binned count rates, where the fluctuations are expected to be averaged out.

	\item {\bf SPOL} \\
		The SPOL CCD Imaging/Spectropolarimeter at the Steward Observatory at the University of Arizona \citep{CCD-SPOL} regularly monitors Mkn 421 as a part of the {\it Fermi} multiwavelength support program. The publicly available optical R-band, and V-band photometric and linear polarization data are downloaded from SPOL website\footnote{http://james.as.arizona.edu/$\sim$psmith/Fermi/}.

	\item {\bf OVRO} \\
		As a part of the Fermi monitoring program, the Owens Valley Radio Observatory (OVRO) \citep{ovro} has been regularly observing this source since 2007. Flux measurements at 15 GHZ are taken directly from their website\footnote{http://www.astro.caltech.edu/ovroblazars/}.

\end{enumerate}
 \section[sect:Multi]{Multiwavelength temporal study}\label{sect:Multi}

\begin{figure*}
        \centering
        \includegraphics[scale=1.4]{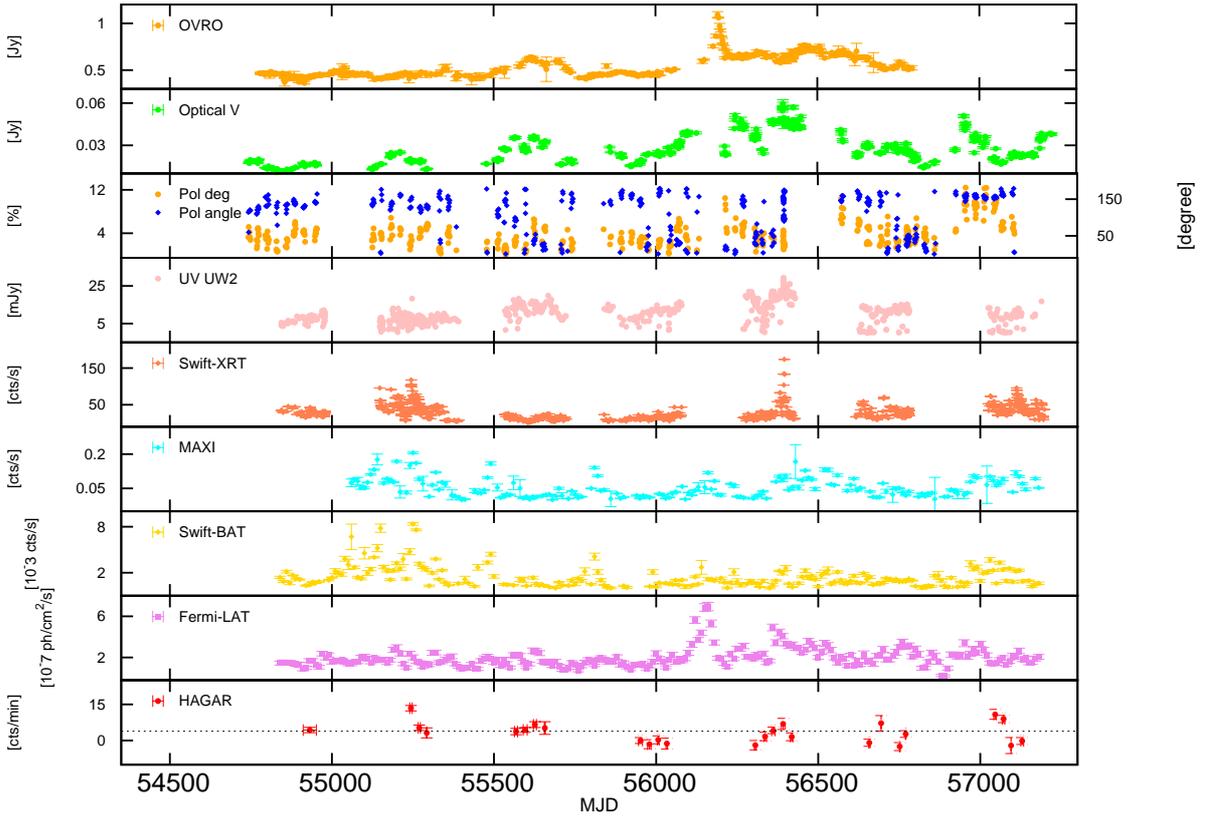}
        \caption{Multiwavelength light curve of Mkn421 from 2009-2015 showing in
		panel 1 : Radio fluxes at 15GHz from the OVRO; panel 2: Optical V-band flux from the CCD-SPOL; panel 3: Degree and angle of the optical polarization using data from CCD-SPOL; panel 4: UV flux (in the Swift-UVOT UW2 band. (Fluxes in the UW1 and WM2 bands follow an exactly similar trend, and thus only one band has been plotted to avoid cluttering); panel 5: {\it Swift}-XRT count rates; panel 6: MAXI count rate (averaged over 10 days); panel 7 :{\it Swift}-BAT count rates (10 days averaged); panel 8:{\it Fermi}-LAT flux in $10^{-7} ph/cm^2/sec$ averaged over 10 days; panel 9: HAGAR counts/min, averaged over each observation season. The black dotted line denotes the Crab count rate.}
        \label{fig:lc_all}
\end{figure*}

 The multiwavelength lightcurve of Mkn 421 from all available instruments during 2009-2015 is given in Fig. \ref{fig:lc_all}. The integration time for each panel is chosen according to the instrument sensitivities. HAGAR points are averaged over each observation season ($\sim 10$ days), likewise the Fermi-LAT, Swift-BAT and MAXI points are also averaged over $10$ days. Each Swift-XRT and Swift-UVOT point corresponds to one pointing ($\sim$ hours) and ($\sim$ minutes) respectively. CCD-SPOL and OVRO points have integration times of a few seconds. As can be clearly seen, the source is variable across all wavelengths, and many flares can be identified. The most dramatic flare in the GeV band occurred during in 2012 (MJD $\sim 56140 - 56180$; Calendar dates 2012-08-01 to 2012-09-10), which is reflected in the radio band after a gap of one and half months (MJD $\sim 56193$, Calendar date 2012-09-23). In contrast, the largest X-ray/Optical flare happened during April 2013, which has been studied in detail by various authors \citep{my421,Pian421,vaidehi421}. In this section, we study the flux correlations between the various frequencies, and quantify the nature of the variability.
 \subsection{Variability and correlations}\label{sect:var_cor}

 We quantify the variability using the fractional variability amplitude parameter $F_{var}$ \citep{Vaughan,varsha_var}. It is calculated as 

\begin{equation} 
        F_{var}=\sqrt{\frac{S^2-\sigma^2_{err}}{\bar{x}^2}}
 ,\end{equation} 
\noindent
where $\sigma^2_{err}$ is the mean square error, $\bar{x}$ the unweighted sample mean, and $S^2$ the sample variance. The error on $F_{var}$ is given as 
\begin{equation}
        \sigma_{F_{var}}= \sqrt{ \left( \sqrt{\frac{1}{2N}}\cdot\frac{\sigma^2_{err}}{\bar{x}^2 F_{var}} \right)^2 +  \left( \sqrt{\frac{\sigma^2_{err}}{N}}\cdot\frac{1}{\bar{x}} \right)^2}
.\end{equation}
\noindent
Here, $N$ is the number of points.

\begin{figure*}
	\centering
	\includegraphics[scale=0.6]{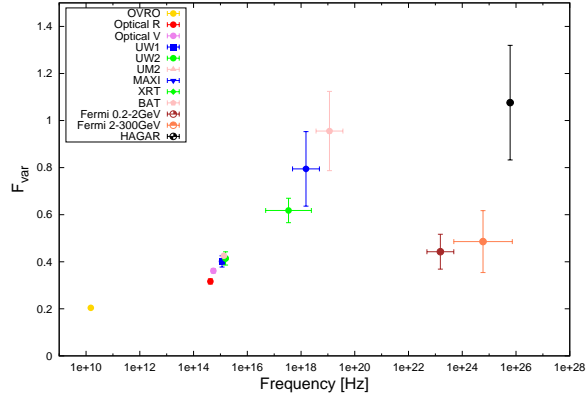}
	\caption{Fractional variability $F_{var}$ as a function of frequency.}
	\label{fig:fvar}
\end{figure*}

The $F_{var}$ values derived from the lightcurves in Fig. \ref{fig:lc_all} are plotted in Fig. \ref{fig:fvar}. To investigate the dependence of variability on the range of timescales, $F_{var}$ was computed with 1day, 10days, 30days and 3months binning (Table \ref{tab:fvar}). Due to sensitivity issues, $F_{var}$ at daily binning cannot be computed for $\gamma$-rays, ie, with the Fermi-LAT and HAGAR. The results remained roughly similar, with the differences showing up mainly in the X-ray bands. For the optical/GeV bands, the difference is $\sim 5\%$. This shows that the low variability seen in the optical/GeV bands is not an artifact of the binning, but is intrinsic to the source. The variability is maximum in the X-ray/VHE bands, $\sim 80\%$, and goes down with frequency. The GeV and UV data show a similar variability index, $\sim 40\%$, which suggests a similar origin for the X-ray and VHE and for the UV and GeV spectrum. Similar results have been reported in \citet{Mkn421_march2010}. 

\begin{figure*}
	\centering
	\subfloat[Swift-XRT]{\includegraphics[scale=0.6]{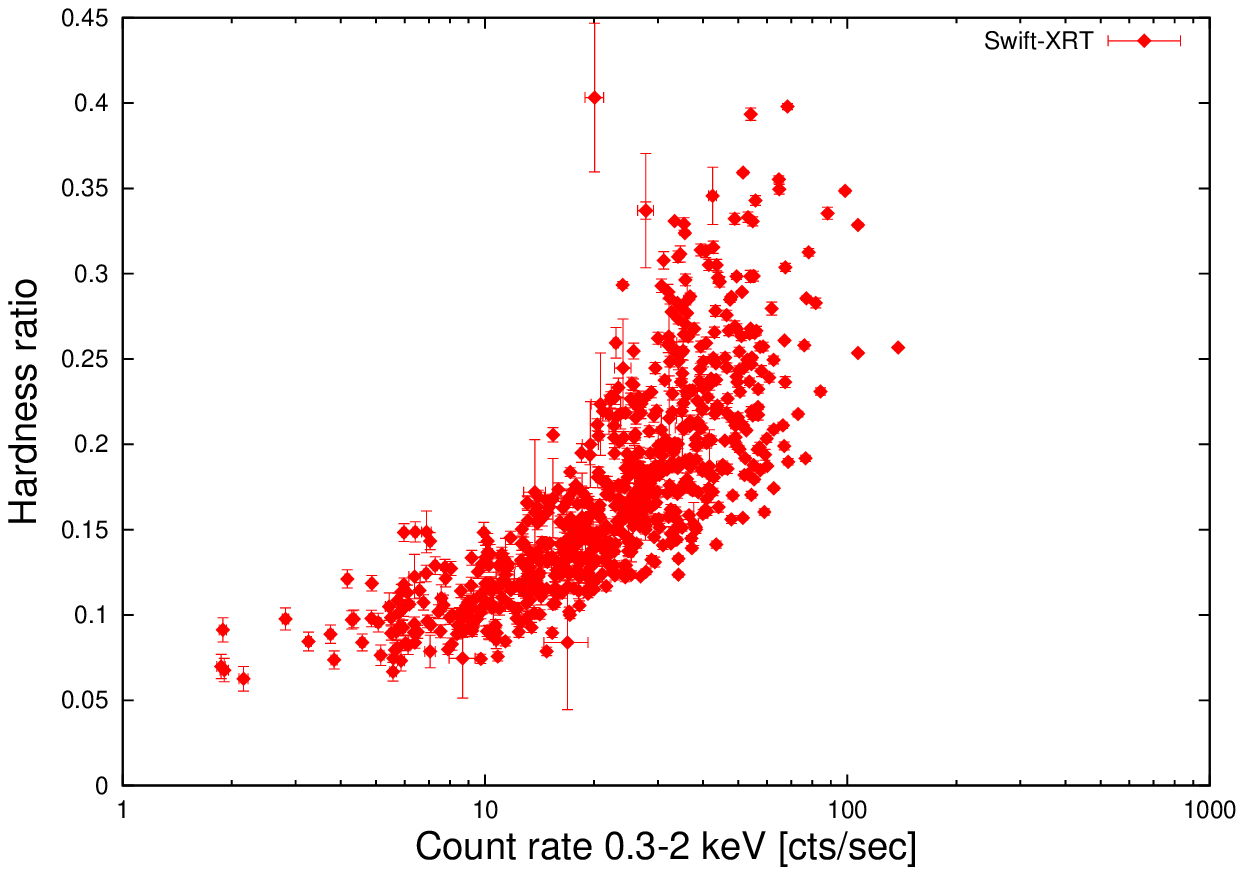}}
      \qquad
      \subfloat[Fermi-LAT]{\includegraphics[scale=0.6]{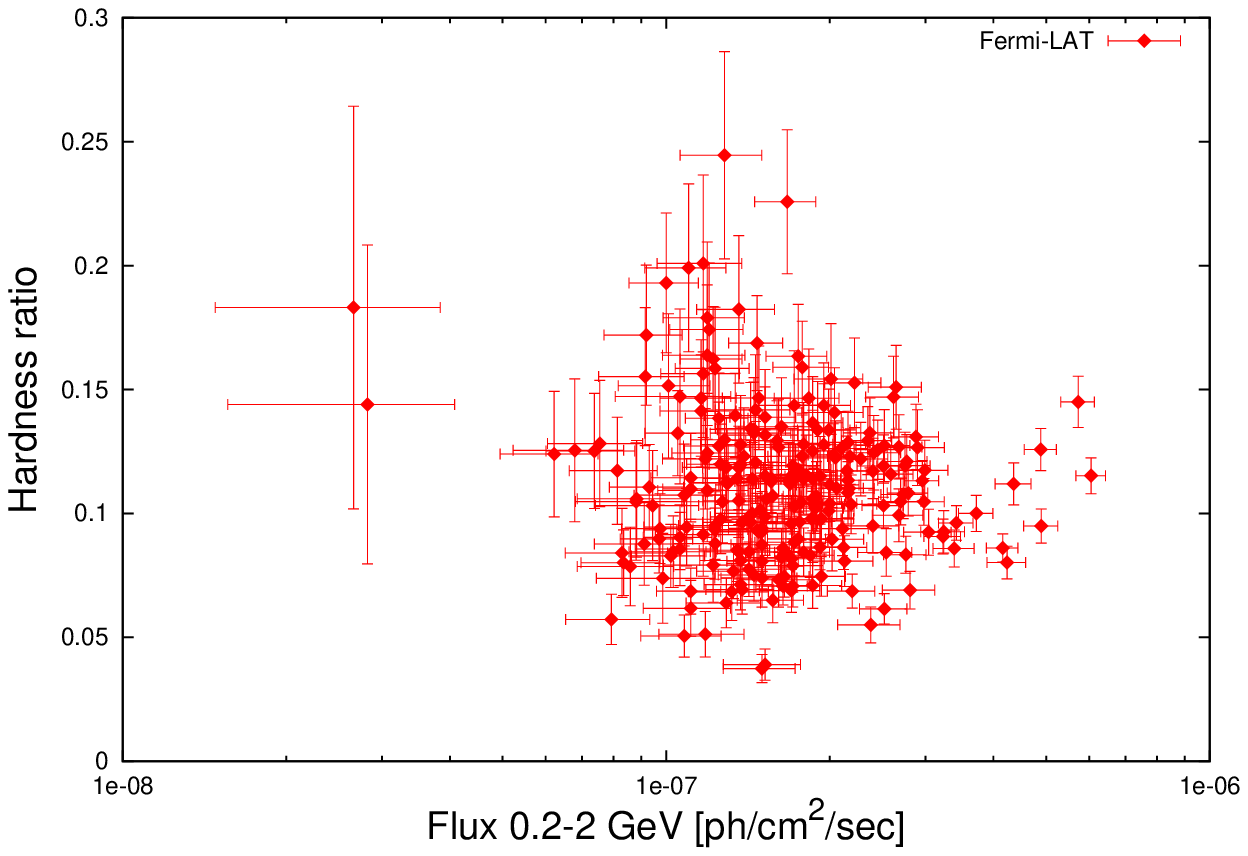}}
	  \caption{Hardness ratio in the X-ray(Swift-XRT) band and GeV (Fermi-LAT) bands. While the trend of spectral hardening with flux is clear in the X-ray band, no such trend is seen in the GeV band.}
	  \label{hardness}
  \end{figure*}

We compute the hardness ratios from Swift-XRT data, defined as the ratio of the flux in the $2-10$ keV band to the flux in $0.3-2$ keV band. As can be seen in Fig. \ref{hardness}, while there is clear trend of spectral hardening with flux seen in the X-ray band ($\rho = 0.83$, prob = $2.1 \times 10^{-16}$), no such evidence is seen in the $\gamma$-ray band ($\rho=0.06$ prob=0.33), again suggesting different origin for the X-ray and GeV data.

\begin{figure*}
	\centering
	\subfloat[Low and high energy X-rays (blue), and low and high energy bands in Fermi (red)]{\includegraphics[scale=0.5]{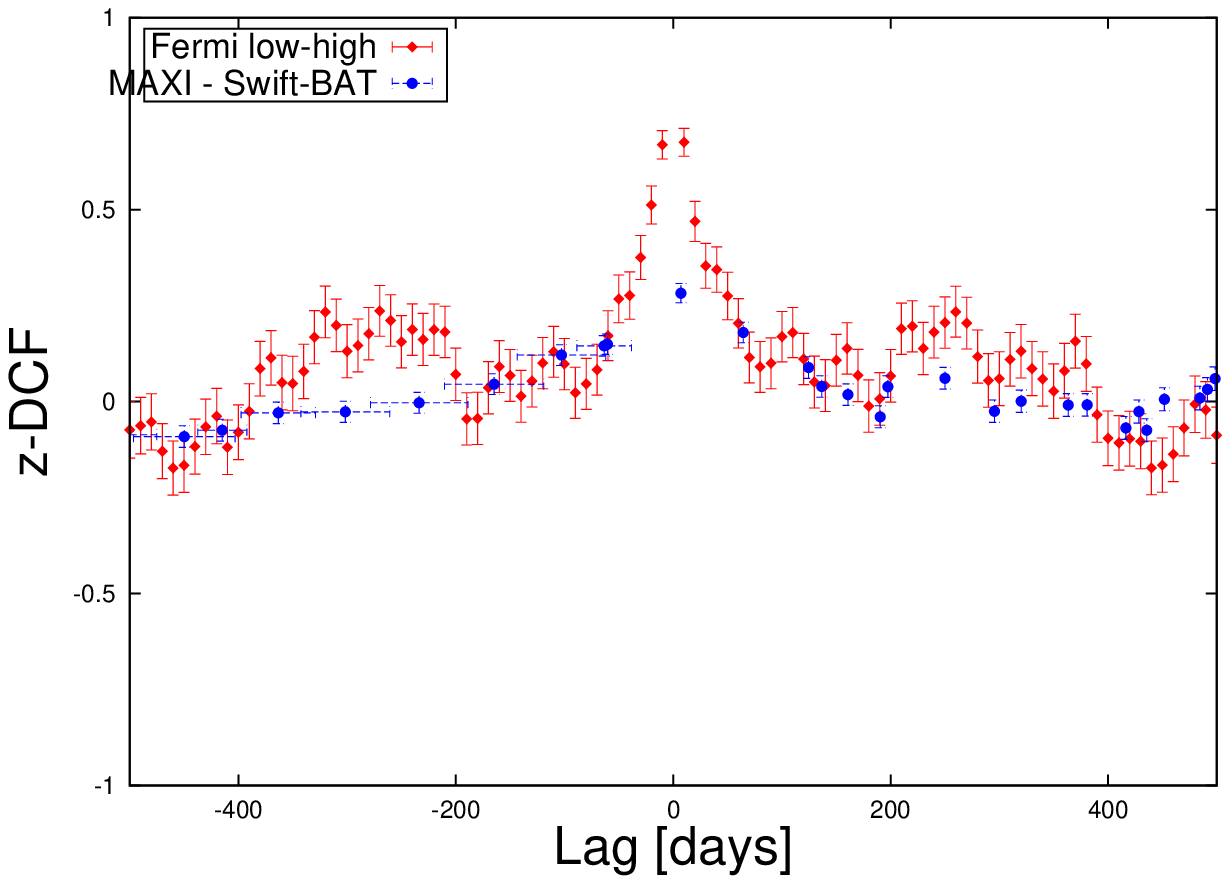}\label{low-high}}
      \qquad
	  \subfloat[Swift-BAT and Fermi-LAT (blue), Swift-XRT and Fermi-LAT (red)]{\includegraphics[scale=0.5]{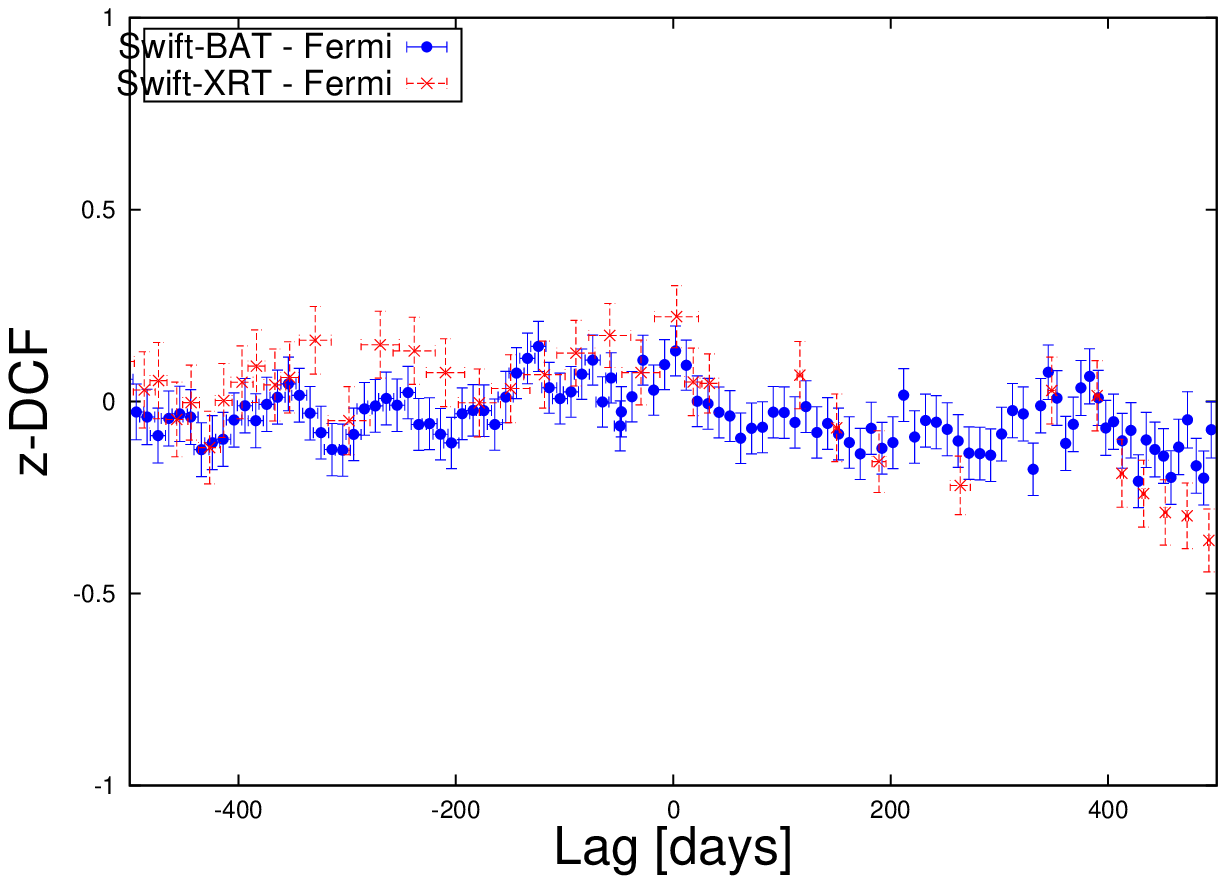}\label{xray-fermi}}
       \qquad
	   \subfloat[Ultra Violet UW2 band - Fermi (blue) and optical V band and Fermi (red) ]{\includegraphics[scale=0.5]{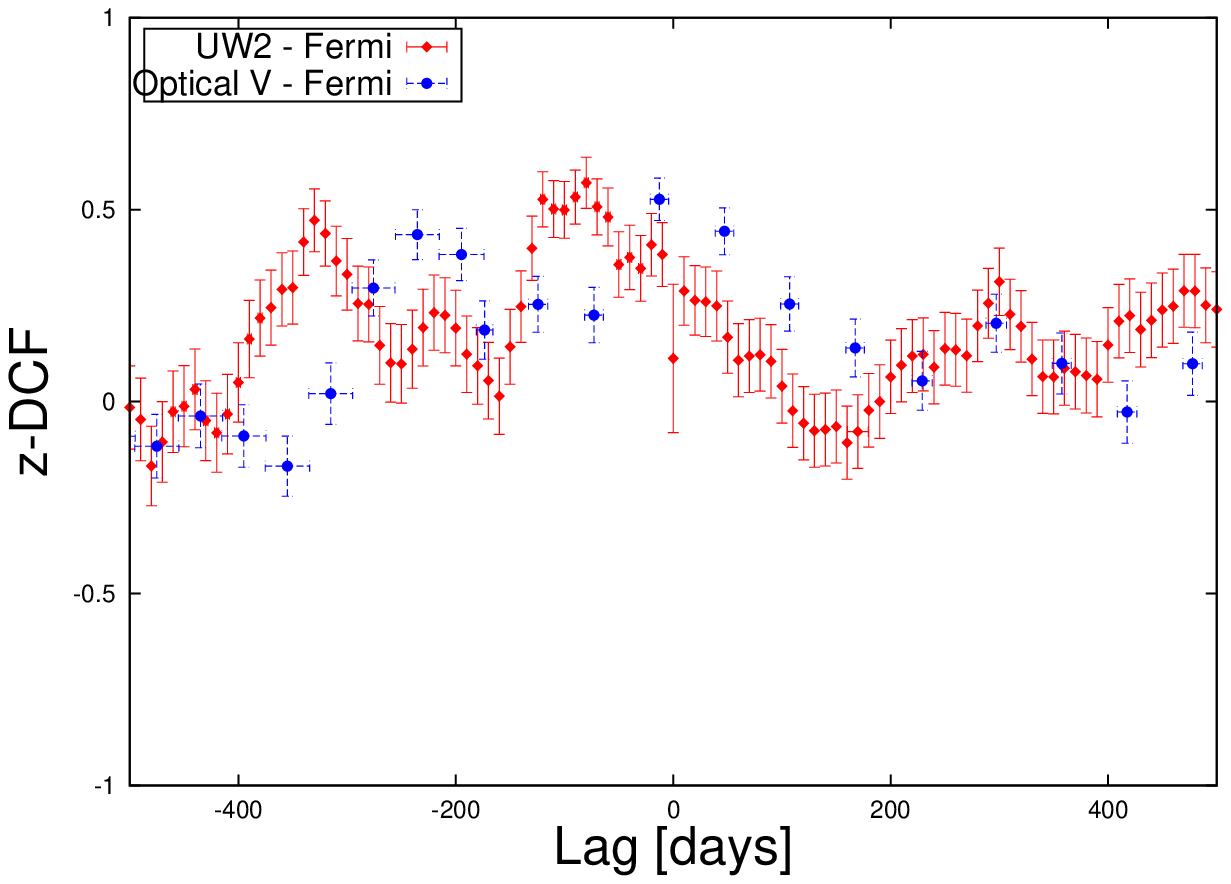}\label{op-fermi}}
      \qquad
	  \subfloat[Radio and Fermi]{\includegraphics[scale=0.5]{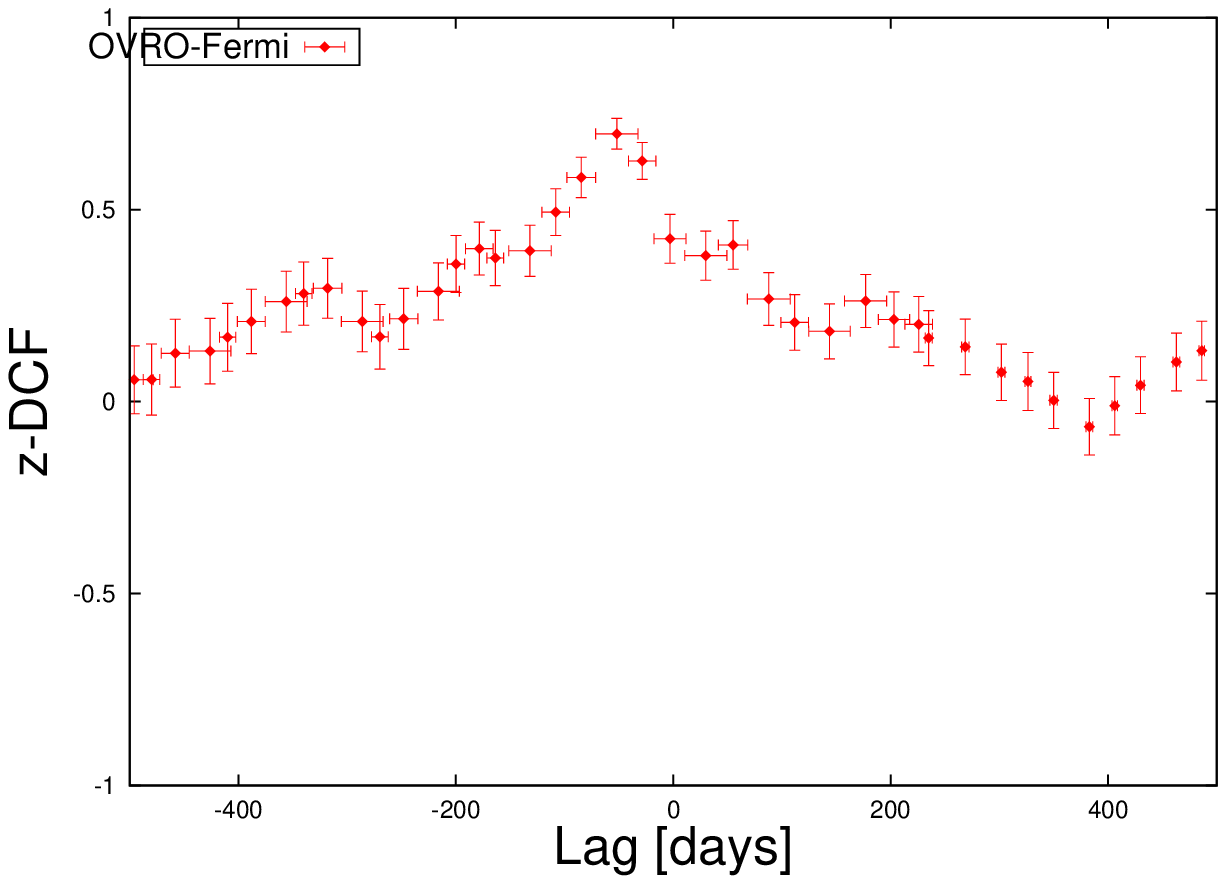}\label{ovro-fermi}}
	\caption{z-DCF between the various wavebands. A positive lag indicates the lower energy flare leads the high energy one.}
	   \label{fig:corr}
\end{figure*}

To study the lags between the various unevenly sampled energy bands, we use the z-transformed discrete cross correlation function, freely available as a FORTRAN 90 routine, with the details of the method described in \citet{zdcf}. A minimum of 11 points are taken in each bin, and points with zero lag are omitted. Since many of the HAGAR seasonal points correspond to upper limits, we avoid the use of HAGAR data in detailed temporal studies. The computed z-DCFs are plotted in Fig. \ref{fig:corr}. A lag of $73 \pm 20 $ days is observed between the Optical V band and the Ultraviolet UW2 band, with a strong correlation of $0.80 \pm 0.04$. The Fermi low (0.2-2 GeV) and high (2-300GeV) energy bands are highly correlated (z-DCF = $0.67 \pm 0.34$) with no visible lag, whereas the low (2-20 keV, MAXI) and high (15-150 keV, Swift-BAT) energy X-ray bands are less strong correlated (z-DCF = $ 0.28 \pm 0.02 $) at a lag of $7.1 \pm 0.5$ days (Fig. \ref{low-high}). There is no correlation between the Fermi and the X-ray bands (Fig. \ref{xray-fermi}). The Fermi flux does not show any lag with the Optical flux (z-DCF = $0.52 \pm 0.06$), but leads the UV flux by $80 \pm 1$ days (z-DCF = $0.57 \pm 0.07$) (Fig. \ref{op-fermi}). Interestingly, there is strong radio-$\gamma$-ray correlation ( z-DCF = $0.69 \pm 0.05$) as seen in Fig. \ref{ovro-fermi}, where the radio flux lags behind the $\gamma$-ray flux by $52 \pm 20 $ days. Such strong radio-$\gamma$ ray correlations due to large dominant flares have been previously seen in long term studies of blazars \citep{LAT-RATAN,Wuradgam}.

\subsection{Lognormality}\label{sect:lognorm}

Lognormality is an important statistical process found in many accreting sources like X-ray binaries \citep{lognorm_xrb}. Lognormal fluxes have fluctuations, that are, on average, proportional to the flux itself, and are indicative of an underlying multiplicative, rather than additive physical process. It has been suggested that a lognormal flux behavior in blazars could be indicative of the accretion disk's variability imprint onto the jet \citep{blazar_var}. This behavior in blazars was first clearly detected in the X-ray regime in BL~Lac \citep{giebels_lognorm}, and has been seen across the entire electromagnetic spectrum in PKS~2155$+$304 \citep{lognorm_2155}.

\begin{figure*}
	\centering
  	\subfloat[OVRO]{\includegraphics[scale=0.3]{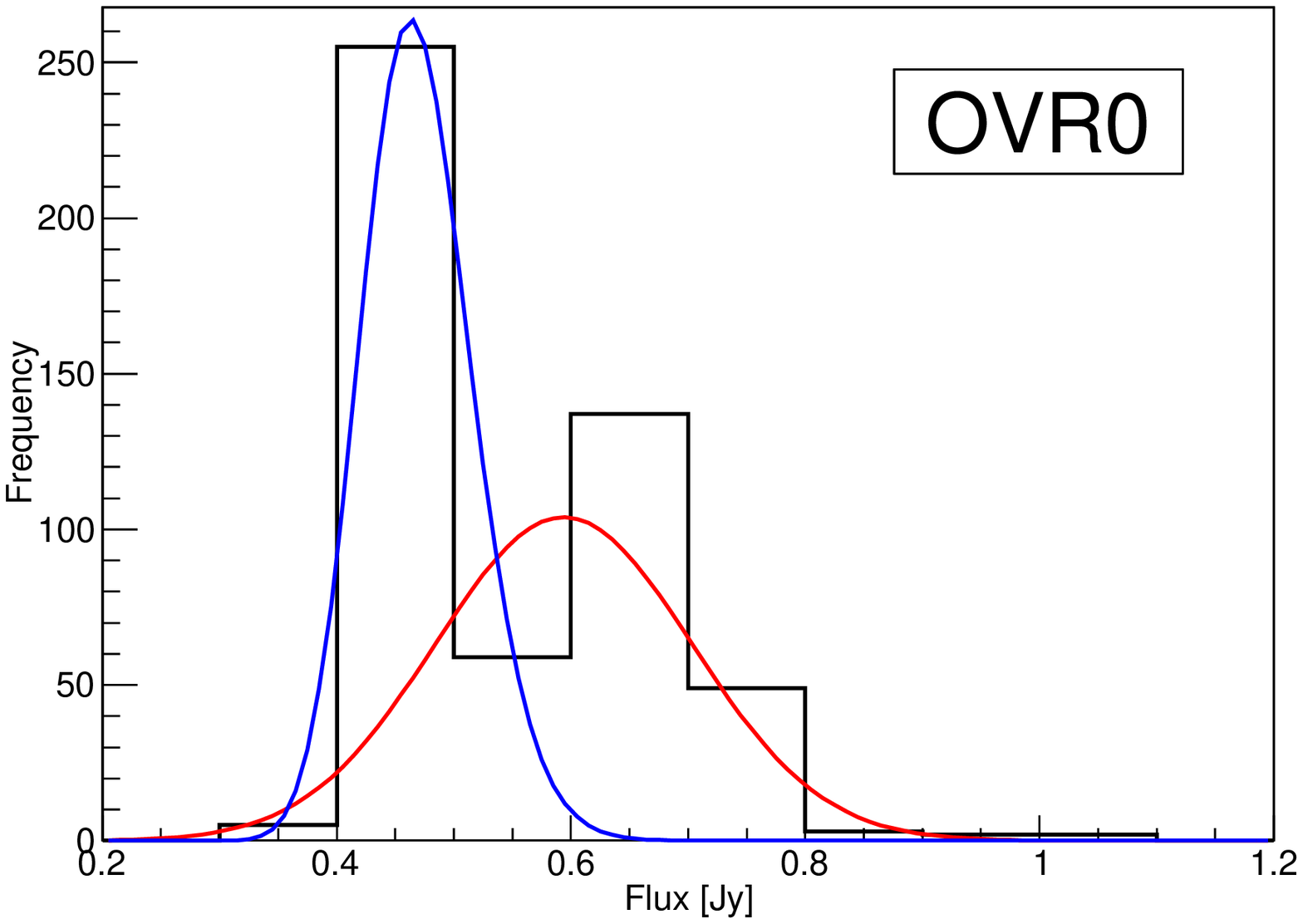}}
      \qquad
      \subfloat[SPOL V band]{\includegraphics[scale=0.3]{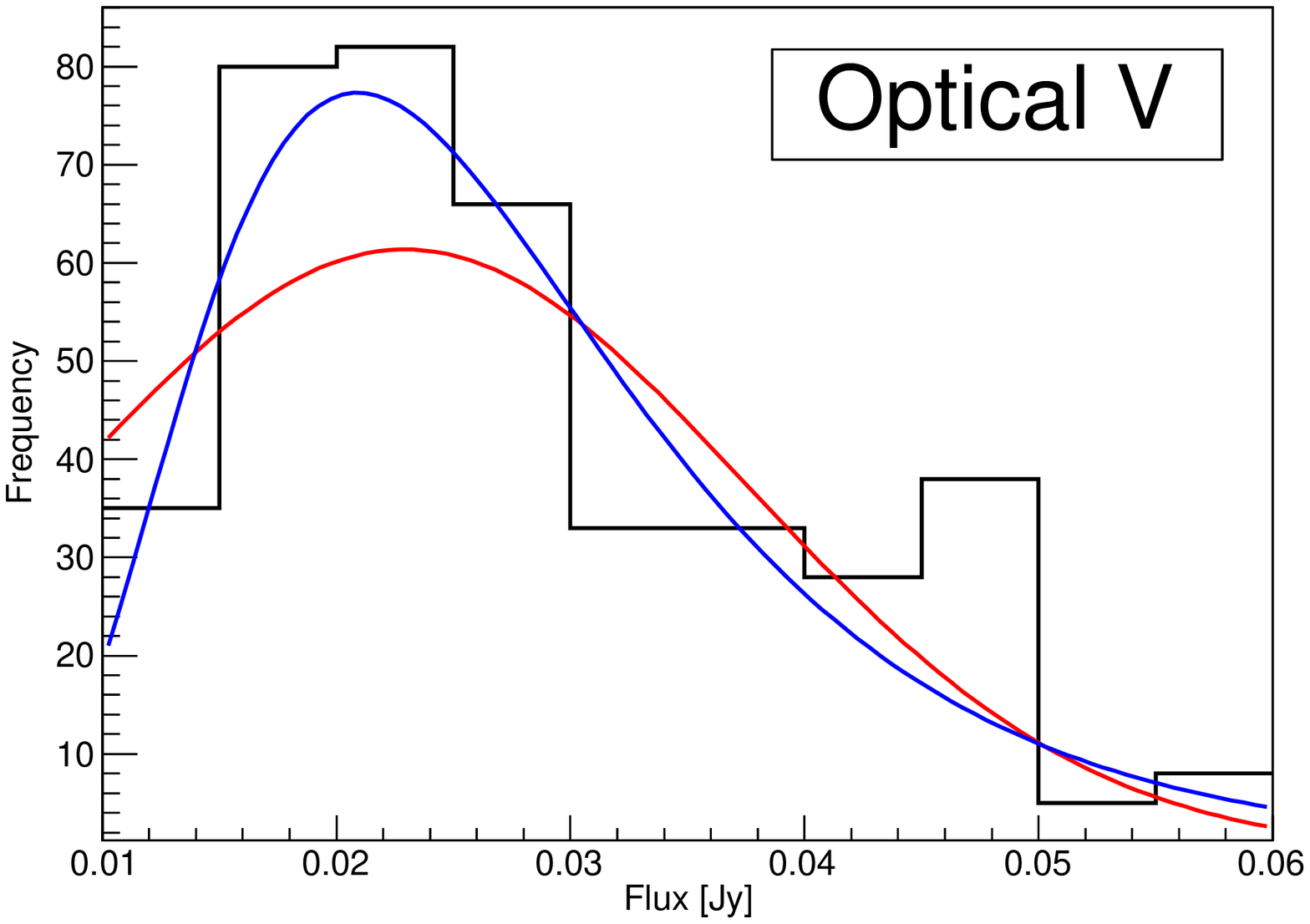}}
       \qquad
	   \subfloat[Swift-UVOT (UW2)]{\includegraphics[scale=0.3]{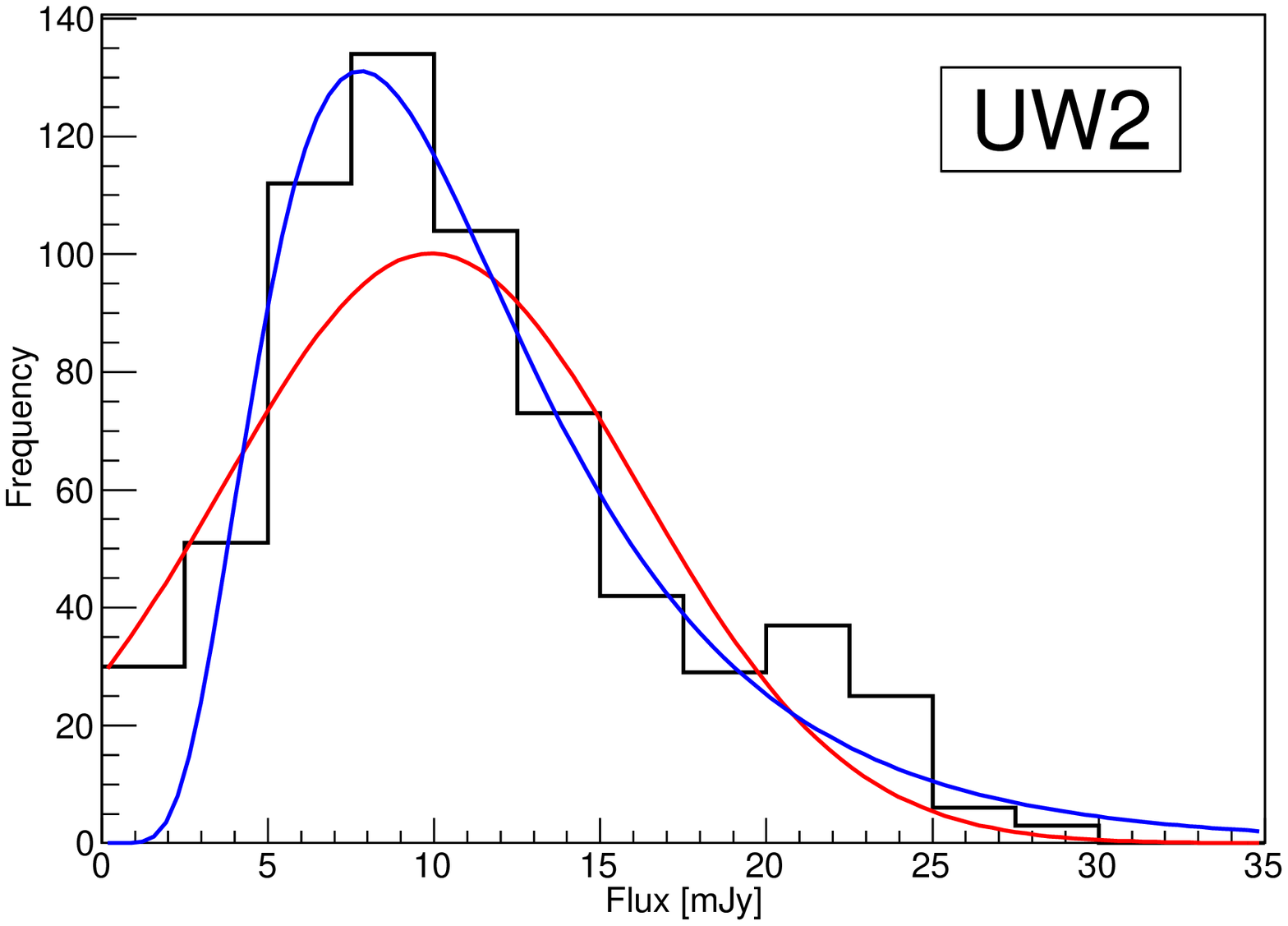}}
      \qquad
	\subfloat[Swift-XRT]{\includegraphics[scale=0.3]{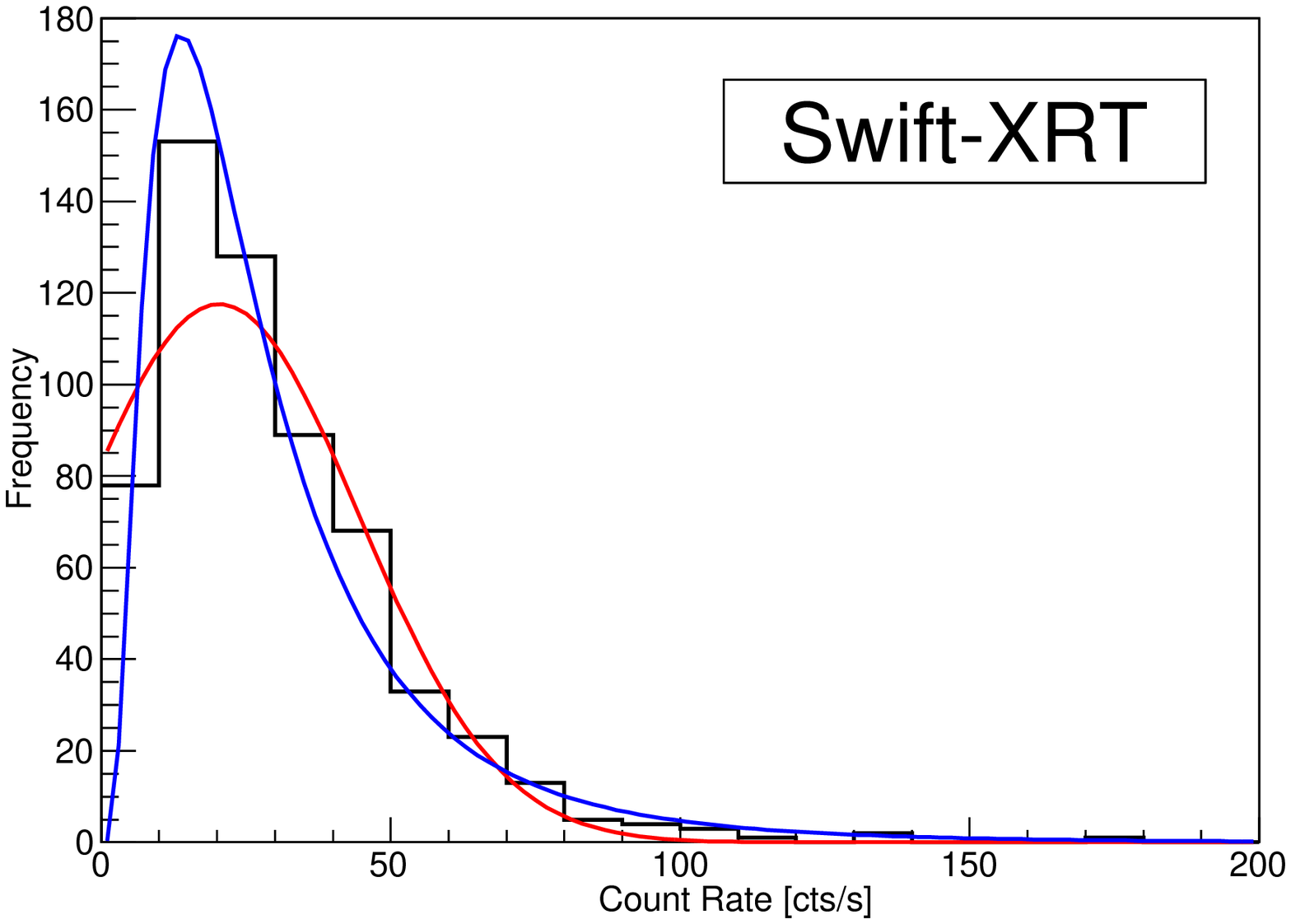}}
       \qquad
	\subfloat[MAXI]{\includegraphics[scale=0.3]{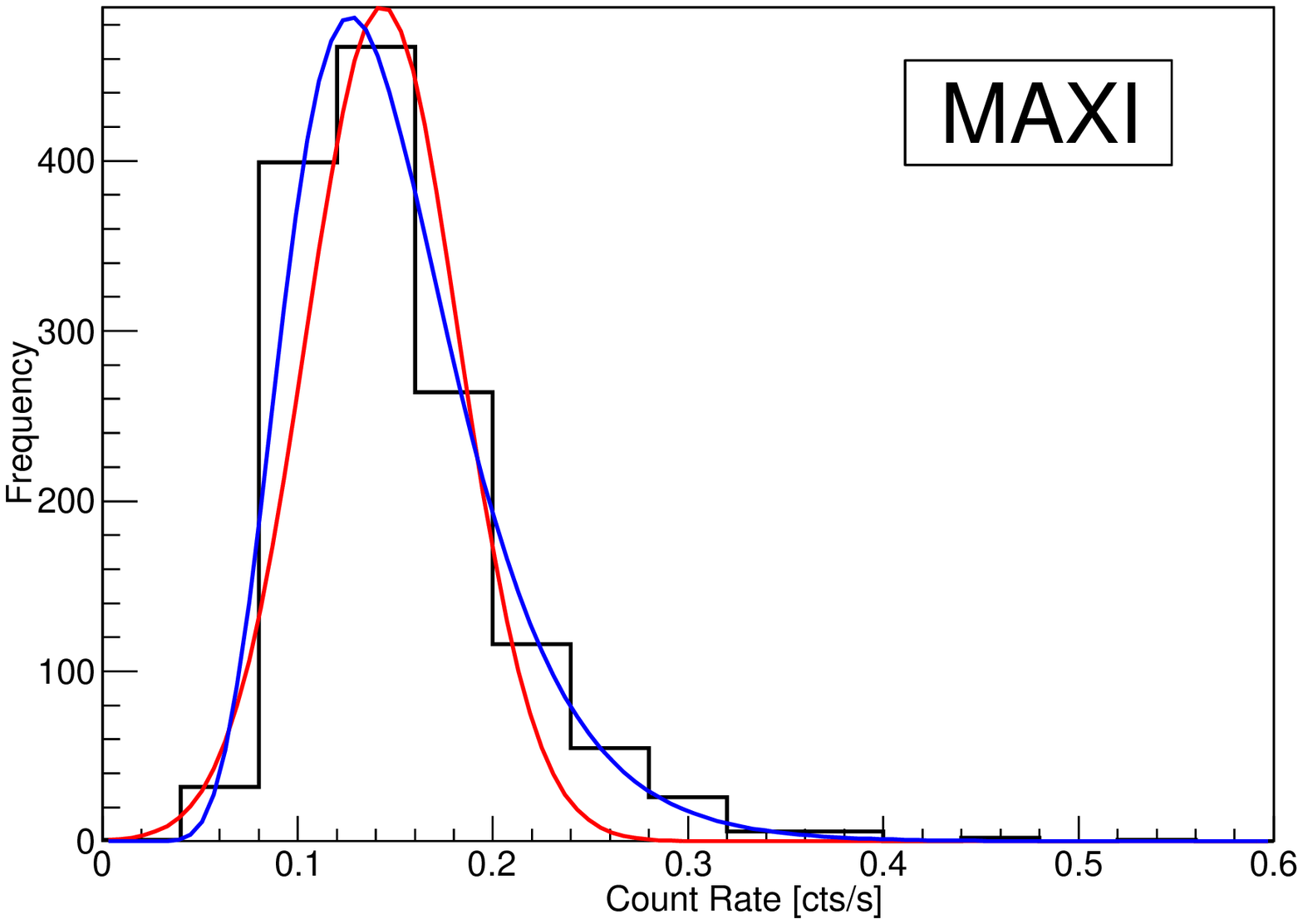}}
       \qquad
       \subfloat[Swift-BAT]{\includegraphics[scale=0.3]{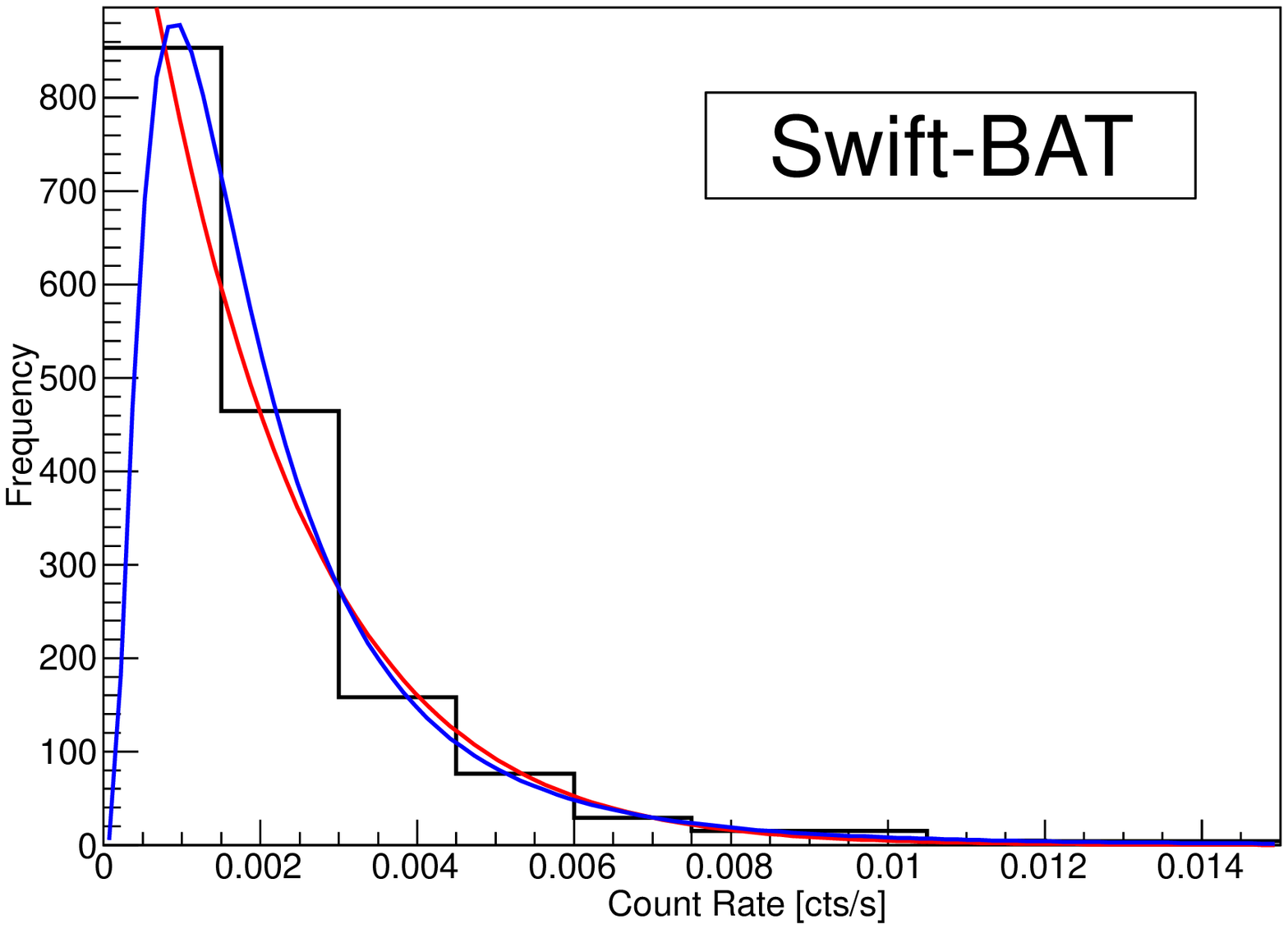}}
       \qquad
	\subfloat[Fermi-LAT]{\includegraphics[scale=0.3]{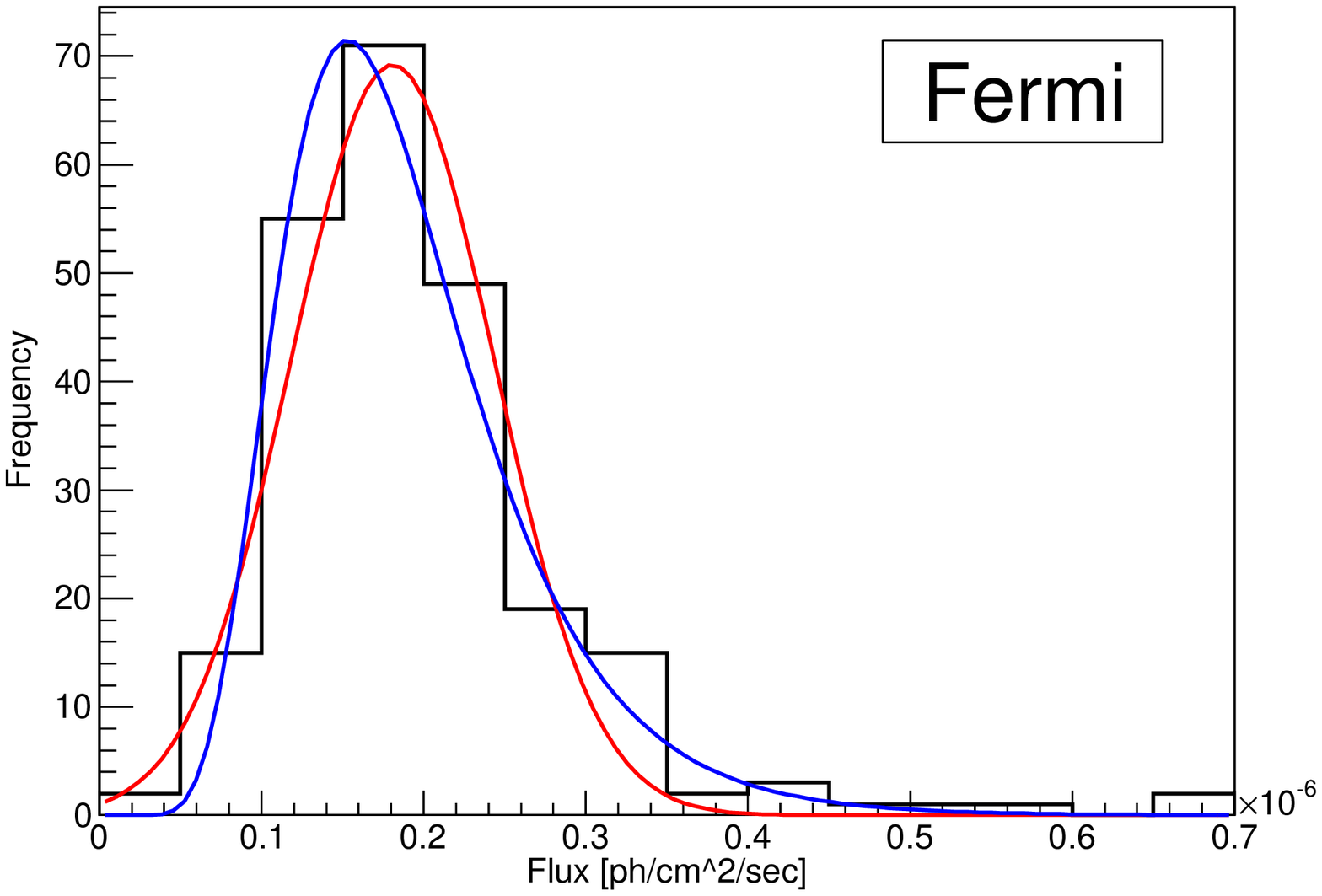}}
	\caption{Histograms of the fluxes (shown in black)  at different wavebands. In all the cases, a lognormal distribution (blue line) fits better than the Gaussian distribution (red line). The reduced chi-squares are given in Table \ref{lognorm}.}
	   \label{Fig:hist}
\end{figure*}

\begin{figure*}
	\centering
  	\subfloat[OVRO]{\includegraphics[scale=0.5]{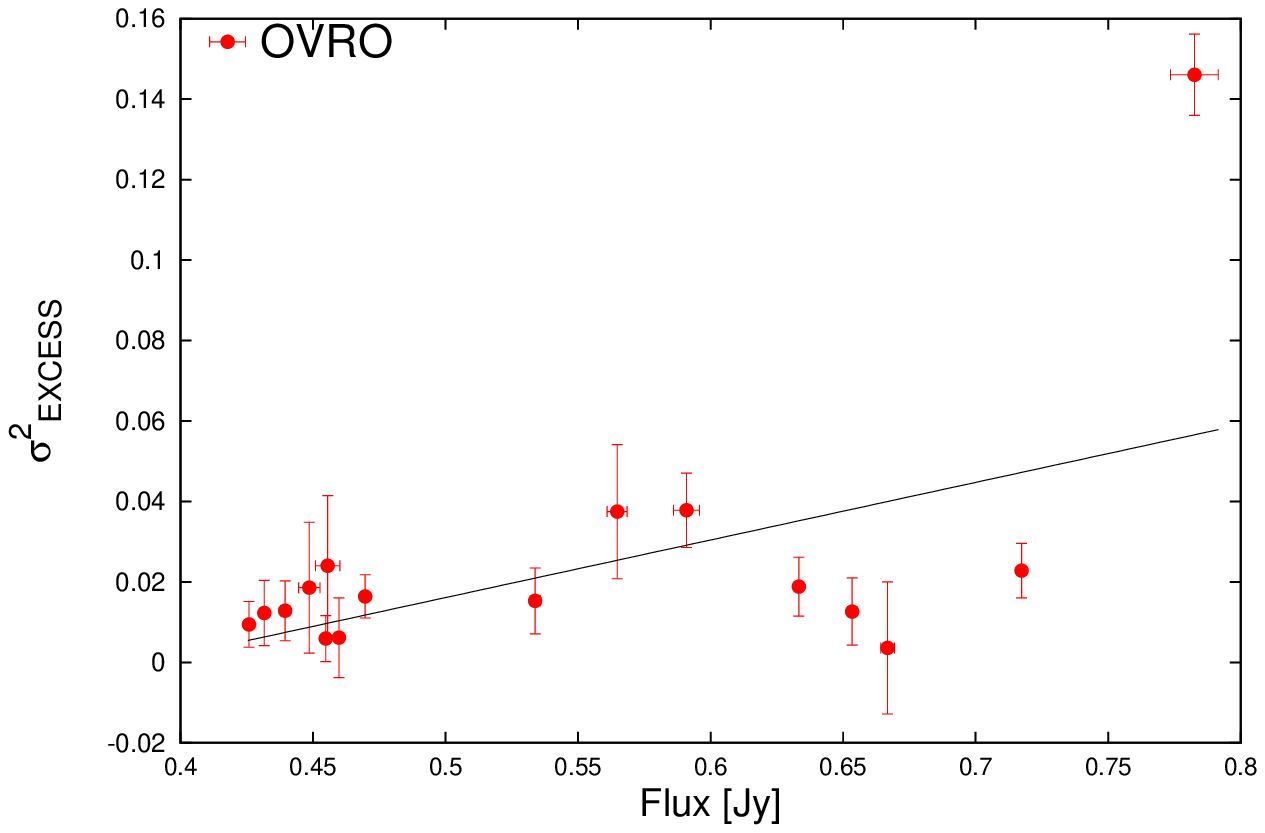}}
      \qquad
      \subfloat[SPOL V band]{\includegraphics[scale=0.5]{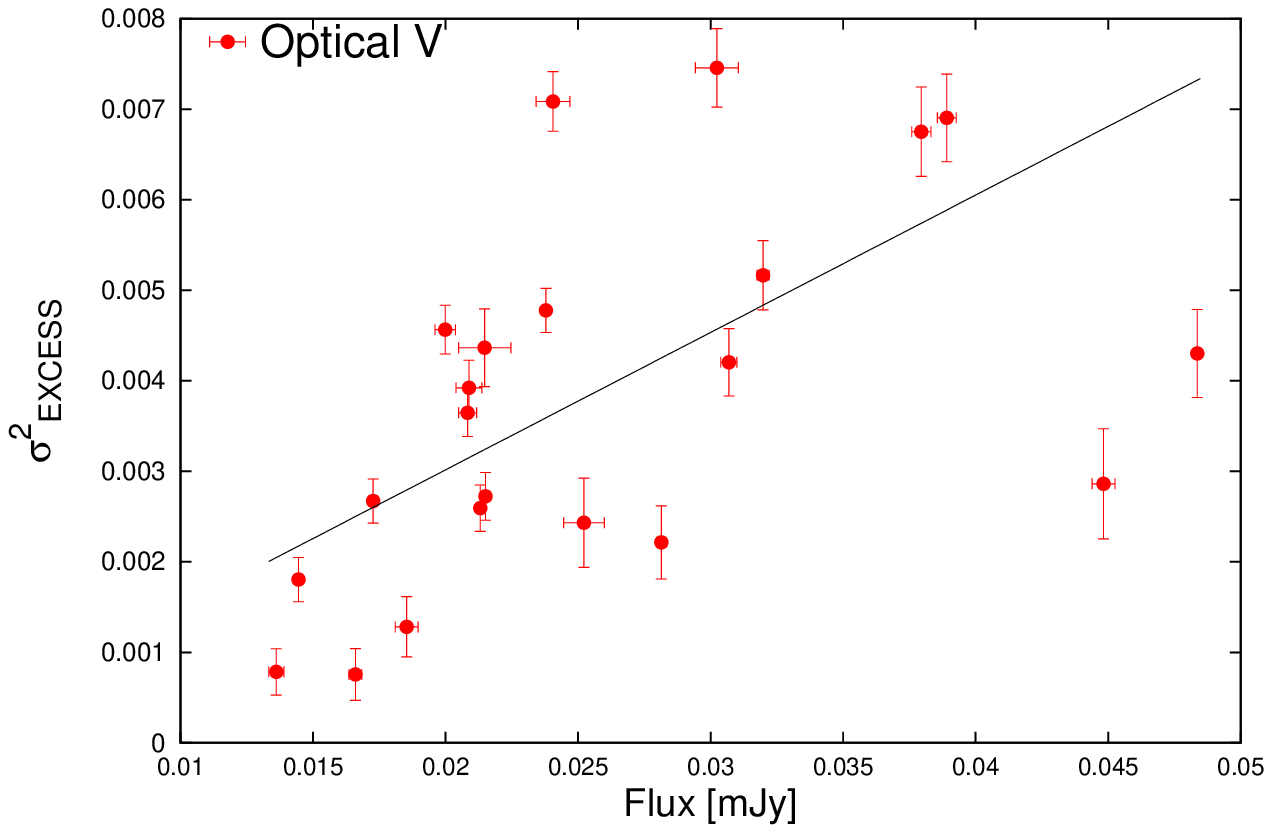}}
       \qquad
	   \subfloat[Swift-UVOT (UW2)]{\includegraphics[scale=0.5]{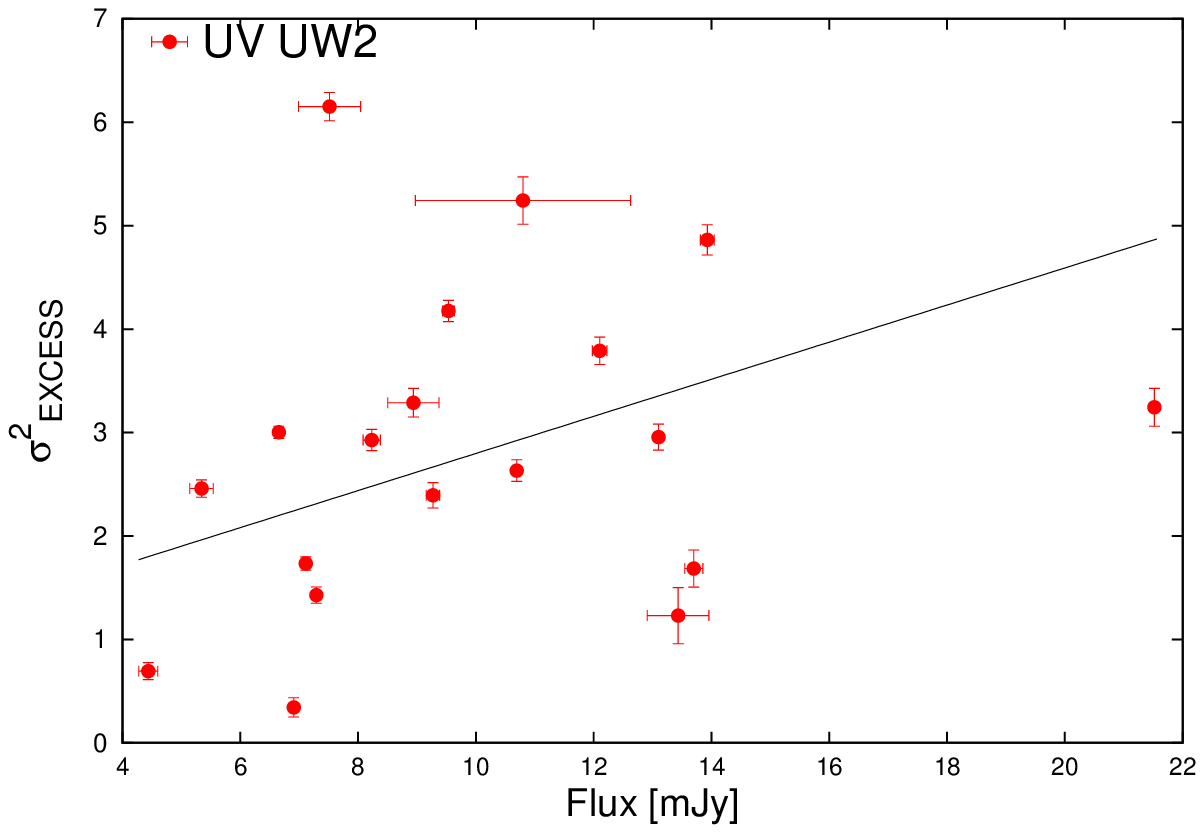}}
      \qquad
	\subfloat[Swift-XRT]{\includegraphics[scale=0.5]{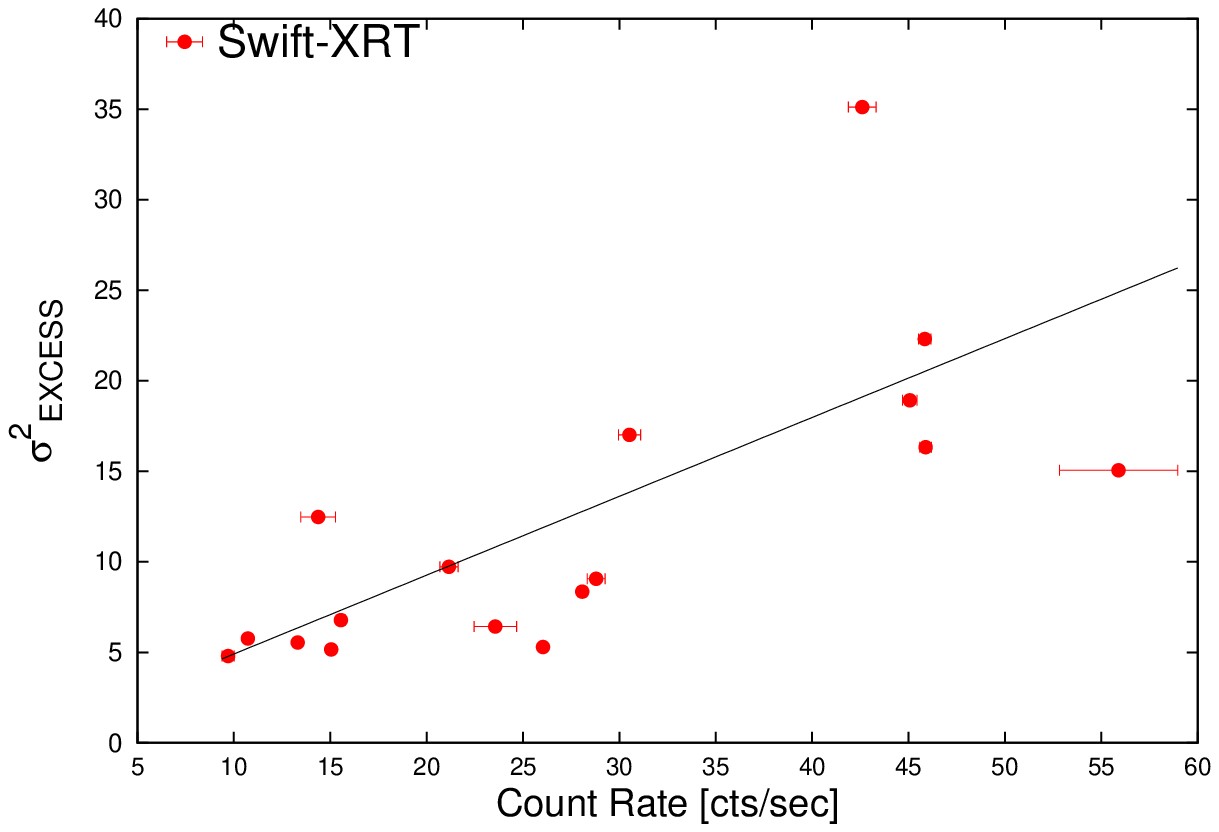}}
       \qquad
	\subfloat[MAXI]{\includegraphics[scale=0.5]{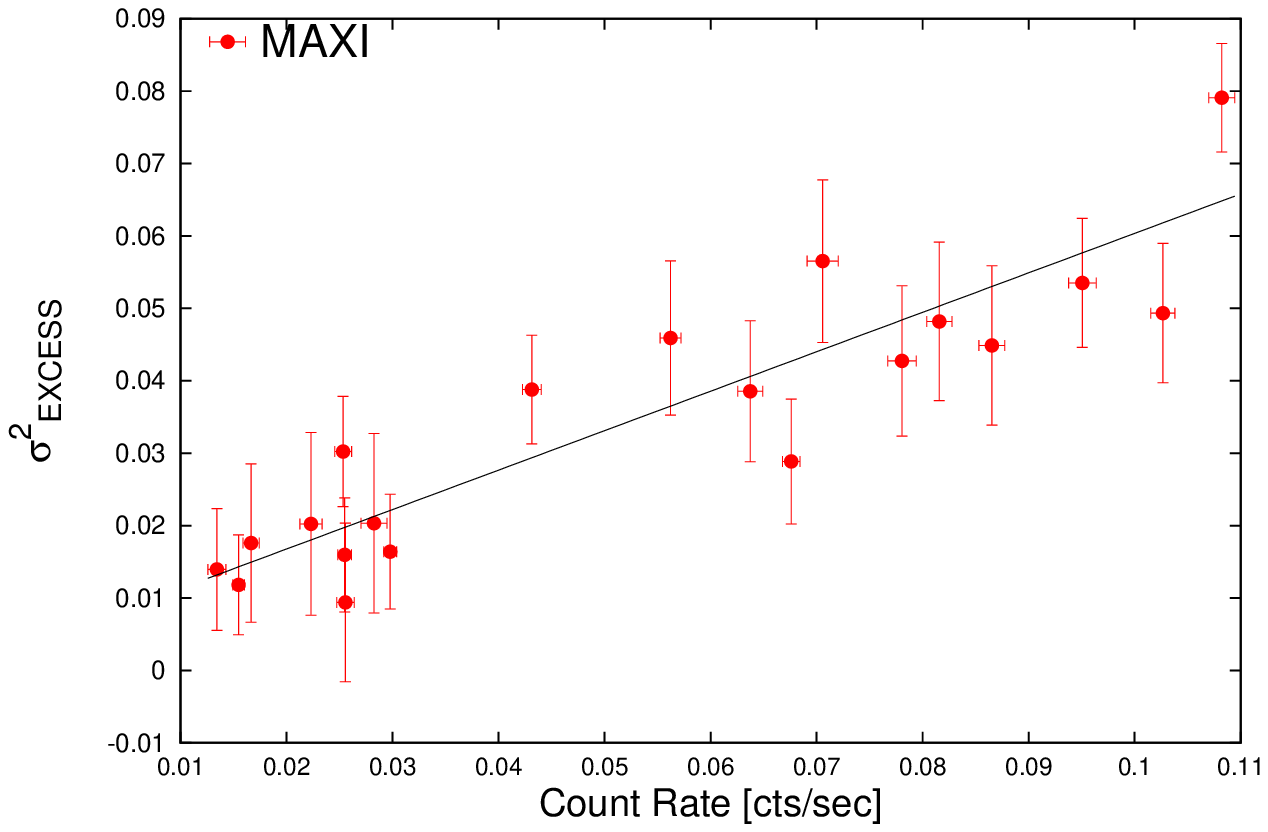}}
       \qquad
       \subfloat[Swift-BAT]{\includegraphics[scale=0.5]{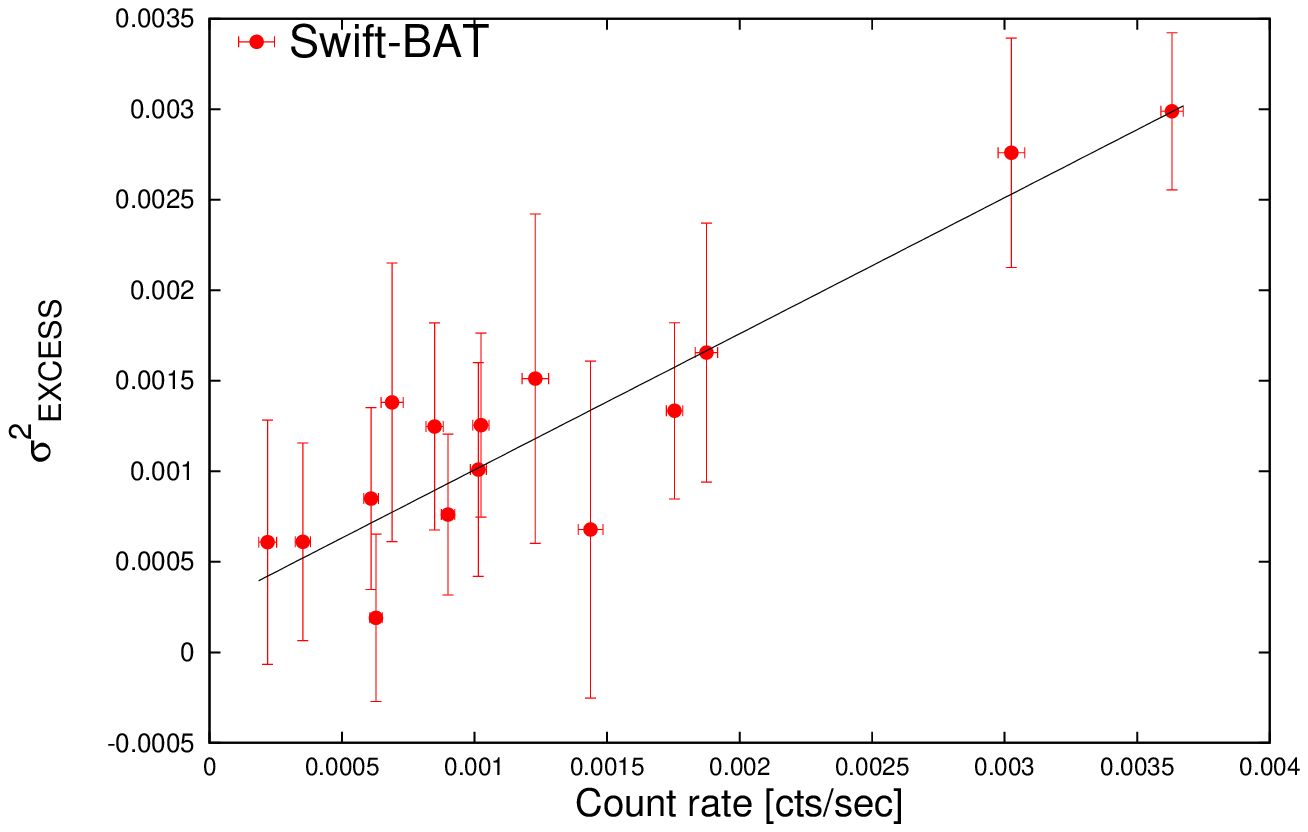}}
       \qquad
	\subfloat[Fermi-LAT]{\includegraphics[scale=0.5]{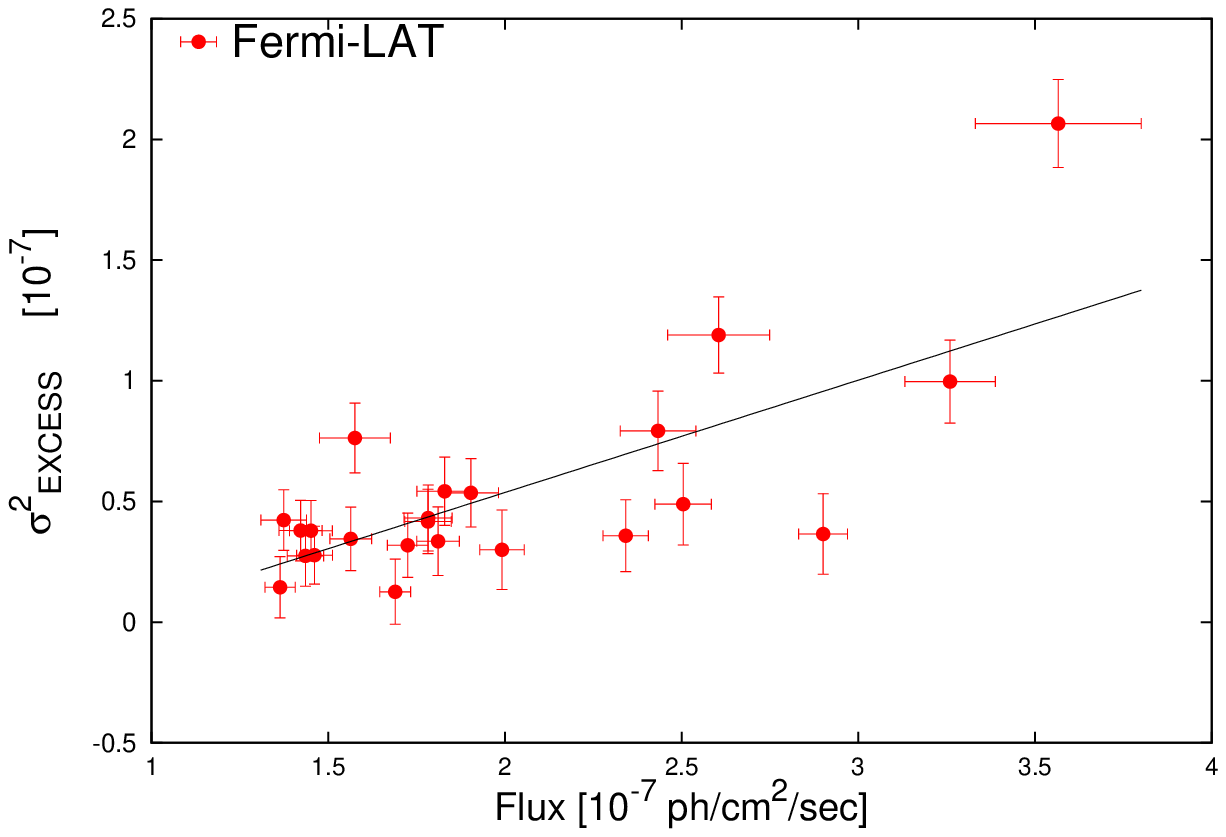}}
       \caption{Excess variance vs the mean flux for the different wavebands. The black lines show the best fit linear regression line.}
	   \label{Fig:excess}
\end{figure*}

To investigate lognormality, we fit the histograms of the observed fluxes with a Gaussian and a Lognormal function (Fig. \ref{Fig:hist}). The reduced chi-squares from the fits are given in Table \ref{lognorm}. The Lognormal function is clearly preferred over the Gaussian, indicating a lognormal trend in the flux distribution. We further plot the excess variance, $\sigma_{EXCESS}= \sqrt{S^2-\sigma^2_{err}}$ vs the mean flux in Fig. \ref{Fig:excess}. Data is binned over a period of 100 days to get sufficient statistics. To test the rms-flux relationship, following \citet{giebels_lognorm, lognorm_2155}, the scatter plot is fit by a constant and a linear function. The linear fit is clearly preferred over the constant, with $r$ as the measure of the correlation coefficient. Spearman's rank correlation ($\rho$) is also computed, the result of which shows that the errors are generally proportional to the flux.
While lognormality is clearly detected for the survey instruments, Fermi-LAT, Swift-BAT and MAXI, the correlations decrease for the other instruments. This is likely due to the fact that the observations from the pointing instruments are biased towards the flaring states. In fact, the histogram of the flux distribution clearly shows two peaks for the OVRO and the R-band data, the second of which can be attributed to the flare states. These flares may be caused by separate ``short-term" phenomena, as compared to the long term flux modulations. This hypothesis is supported by the fact that lognormality is clearly seen by MAXI, but not by Swift-XRT, where the two instruments sample similar energy regimes. Also, the measured radio flux from OVRO is the combined emission from the jet and the radio lobes, the latter of which may contribute to the second component.
	
\section[sect:spectra]{Spectral Modelling}\label{sect:spectra}

Twenty one SEDs are extracted over the past 7 years. We do not bias our SEDs over any flare/quiescent states as in \cite{ybj421}, but choose epochs contemporaneous with HAGAR time periods. During these epochs, the total bolometric luminosity varied almost by a factor of $10$. The broadband emission is assumed to originate from a single spherical zone of radius $R$ filled with a tangled magnetic field $B$. A non-thermal population of electrons, with a broken power law spectral shape,
\begin{equation}
	\label{eq:bknpo}
N (\gamma) d\gamma =\left\{
\begin{array}{ll}
K \gamma^{-p_1}d\gamma,&\mbox {~$\gamma_{min}<\gamma<\gamma_b$~} \\
K \gamma^{(p_2-p_1)}_b \gamma^{-p_2}d\gamma,&\mbox {~$\gamma_b<\gamma<\gamma_{max}$~}
\end{array}
\right.
\end{equation}
\noindent

is assumed to lose its energy through synchrotron and self Compton (SSC) processes. As a result of the relativistic motion of the jet, the radiation is boosted along our line of sight by a Doppler factor $\delta$. This simple model is most commonly used in literature to model the broadband spectrum \citep{GMD,krawz}. Correction for attenuation of TeV photons due to the extragalactic background light (EBL) is accounted for by deconvolving the obtained spectrum with the EBL model of \citep{frans}.
\citet{nijilchi} used the Levenburg-Marquadt algorthim to fit the numerical model to the observed spectrum. We incorporate the numerical SSC model of  \citet{LabSSC} in the XSPEC spectral fitting software to perform a $\chi^2$ minimization. Swift-XRT data are binned to have $\sim 6 $ points (and similarly for NuSTAR) to avoid biasing the fit towards the X-ray energies. The tool {\tt flx2xsp v.2.1} was used to convert the fluxes to pha files for use in XSPEC. Since we are dealing with several different instruments over a broad energy range, model systematics of $5\%$ is assumed.

A good sampling of the entire SED from radio to $\gamma$-rays allows one to perform a reasonable estimation of the physical parameters \citep{GMD,tavKN} in terms of the observed fluxes.  The radius $R$ of the emission region has to be independently estimated from the light travel time argument, $R < c \delta t_{var} / (1+z) $, where $t_{var}$ is the observed flux doubling timescale. However, Mkn 421 shows different flux doubling timescales over different time periods, and since we are dealing mainly with quiescent state emission, we keep $R$ fixed at $6.0 \times 10^{16}$ cm, which was used by \citet{Mkn421_march2010}. This roughly corresponds to a $t_{var} \sim 1$day for $\delta = 21$, which is also frozen during the fitting. The minimum particle Lorentz factor, $\gamma_{min}$ can be fixed at a reasonably low value without affecting the modelling \citep{cerruti_1727}, and is assumed to be  $1000$. The maximum Lorentz factor, $\gamma_{max}$, is kept fixed at $10\gamma_b$.

\begin{figure*}
	  \subfloat[s17]{\includegraphics[scale=0.6]{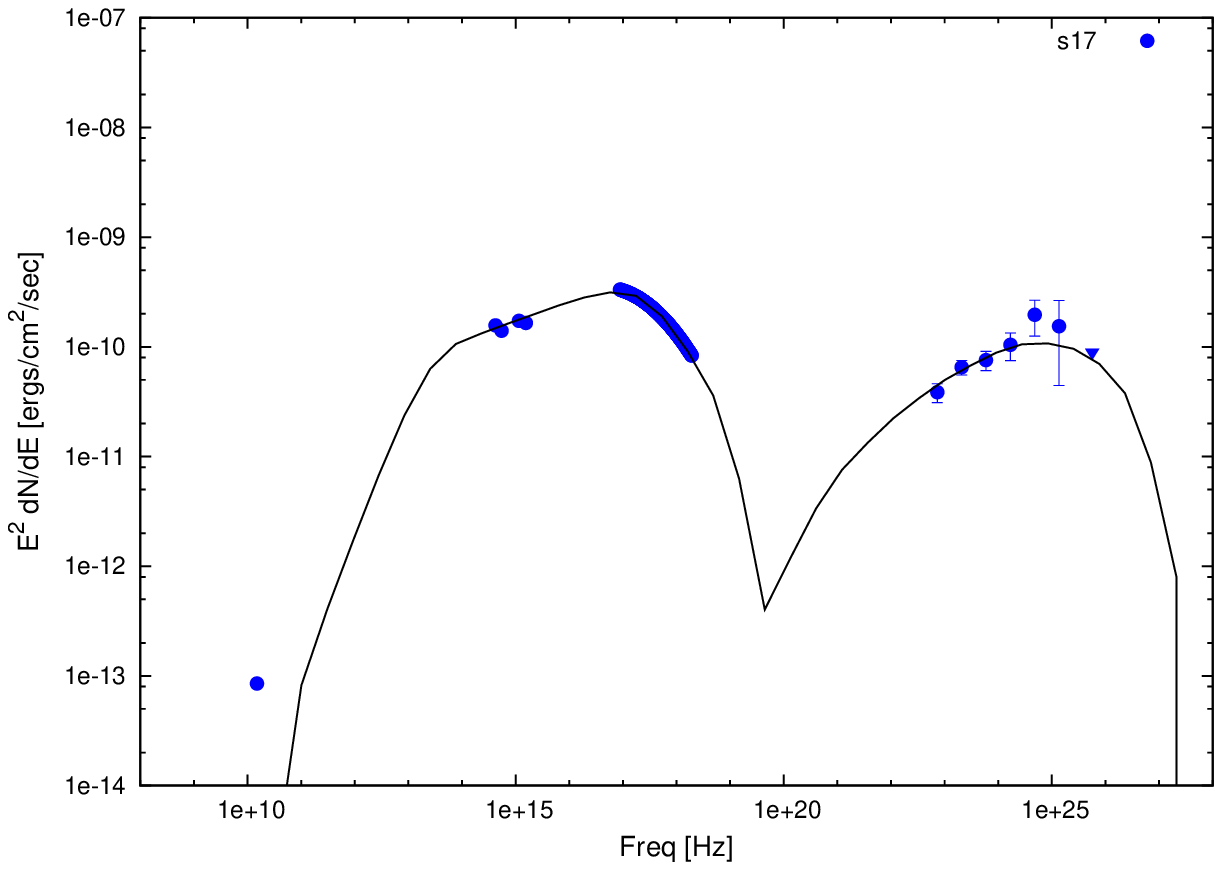}}
      \qquad
  	\subfloat[s18]{\includegraphics[scale=0.6]{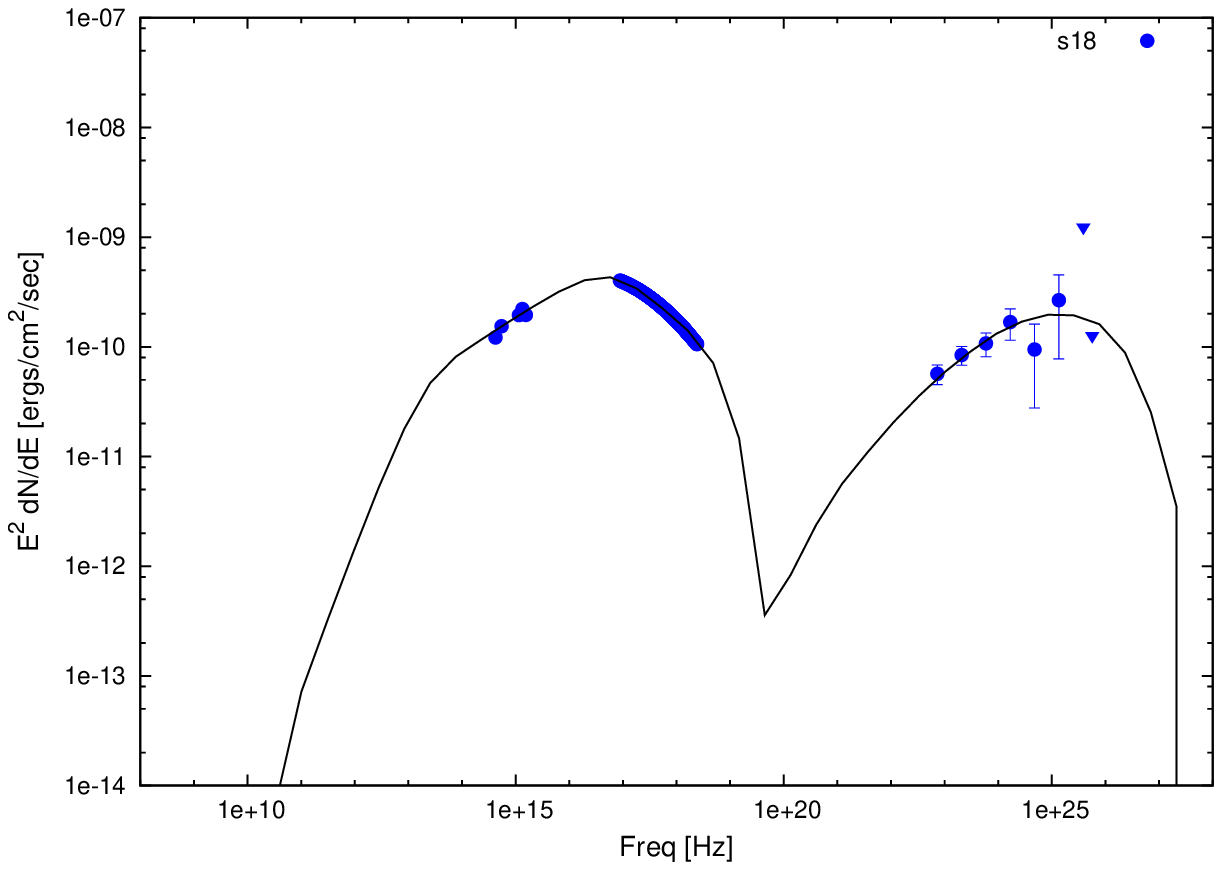}}
      \qquad
  	\subfloat[s19]{\includegraphics[scale=0.6]{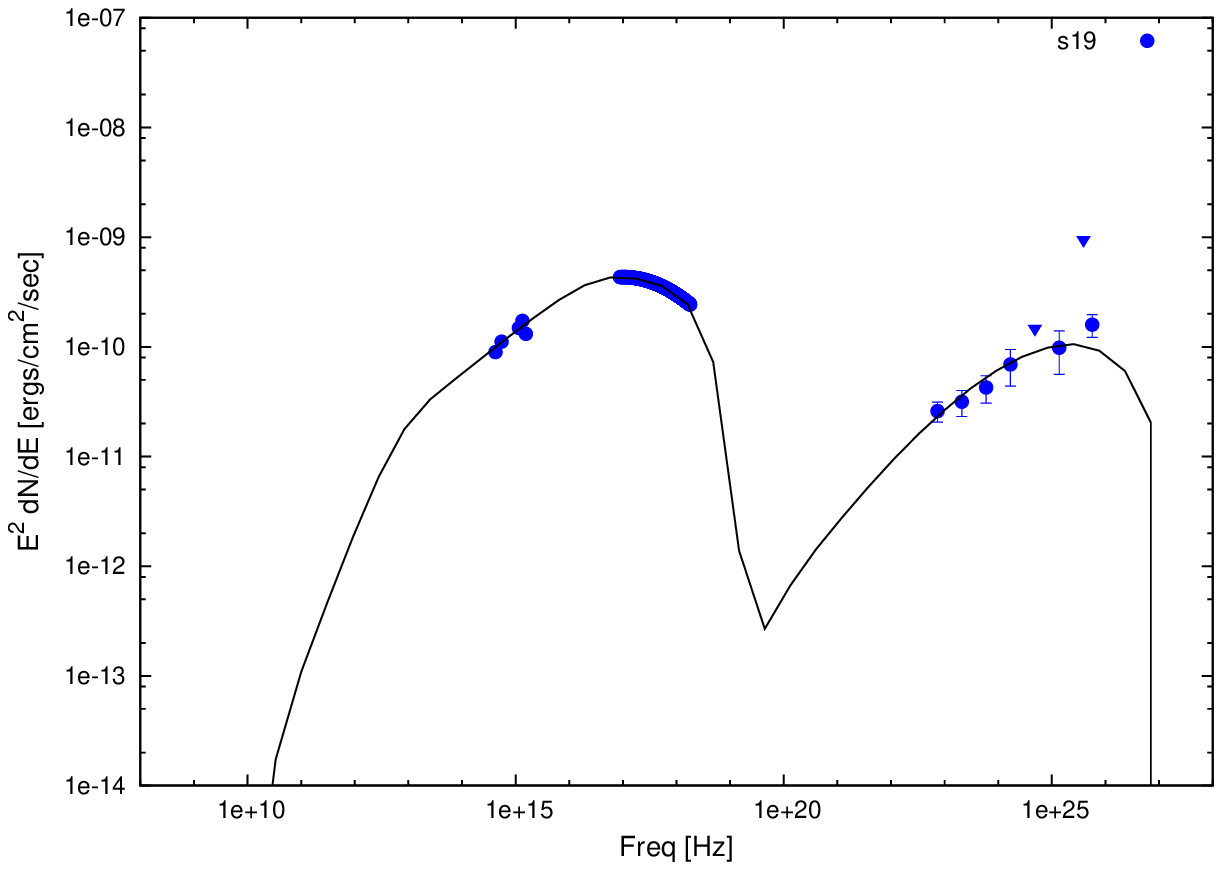}}
      \qquad
	  \subfloat[s20]{\includegraphics[scale=0.6]{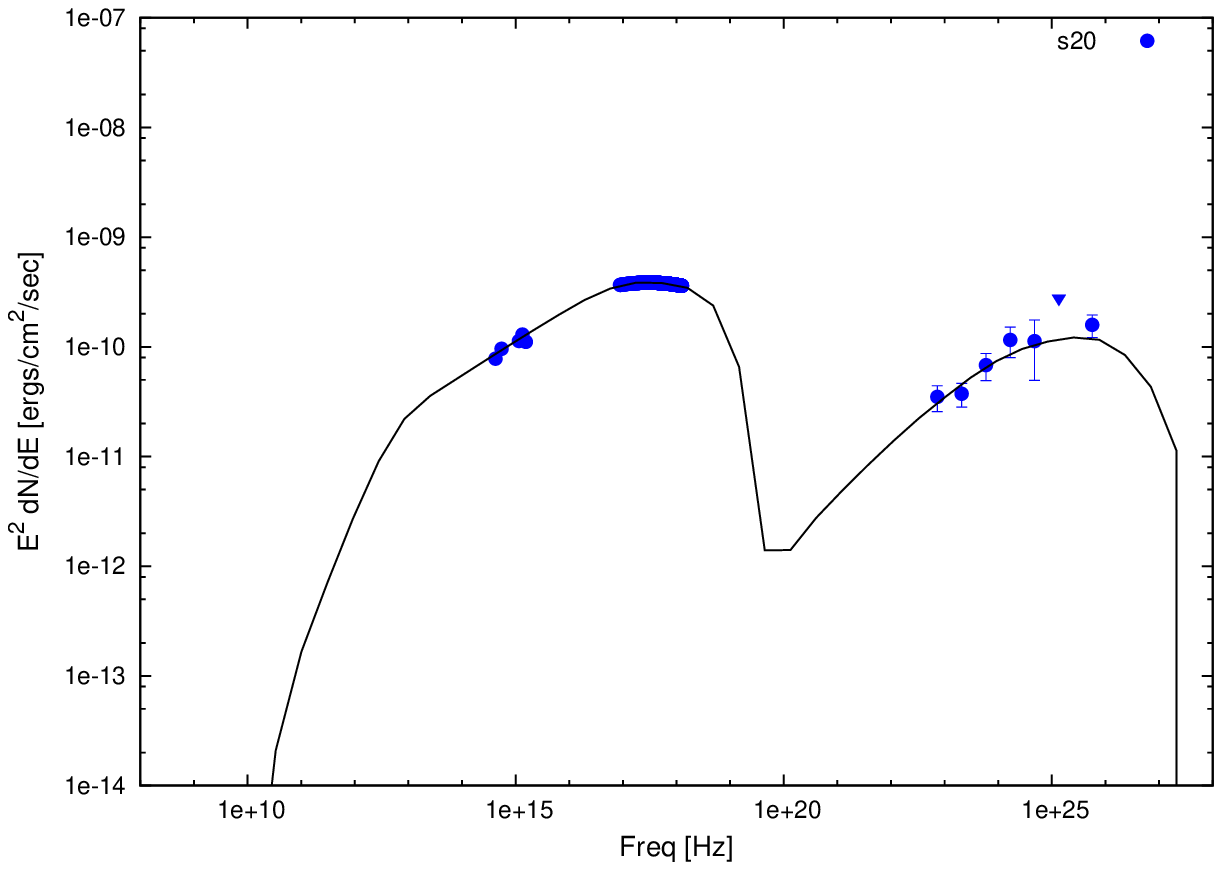}}
      \qquad
  	\subfloat[s21]{\includegraphics[scale=0.6]{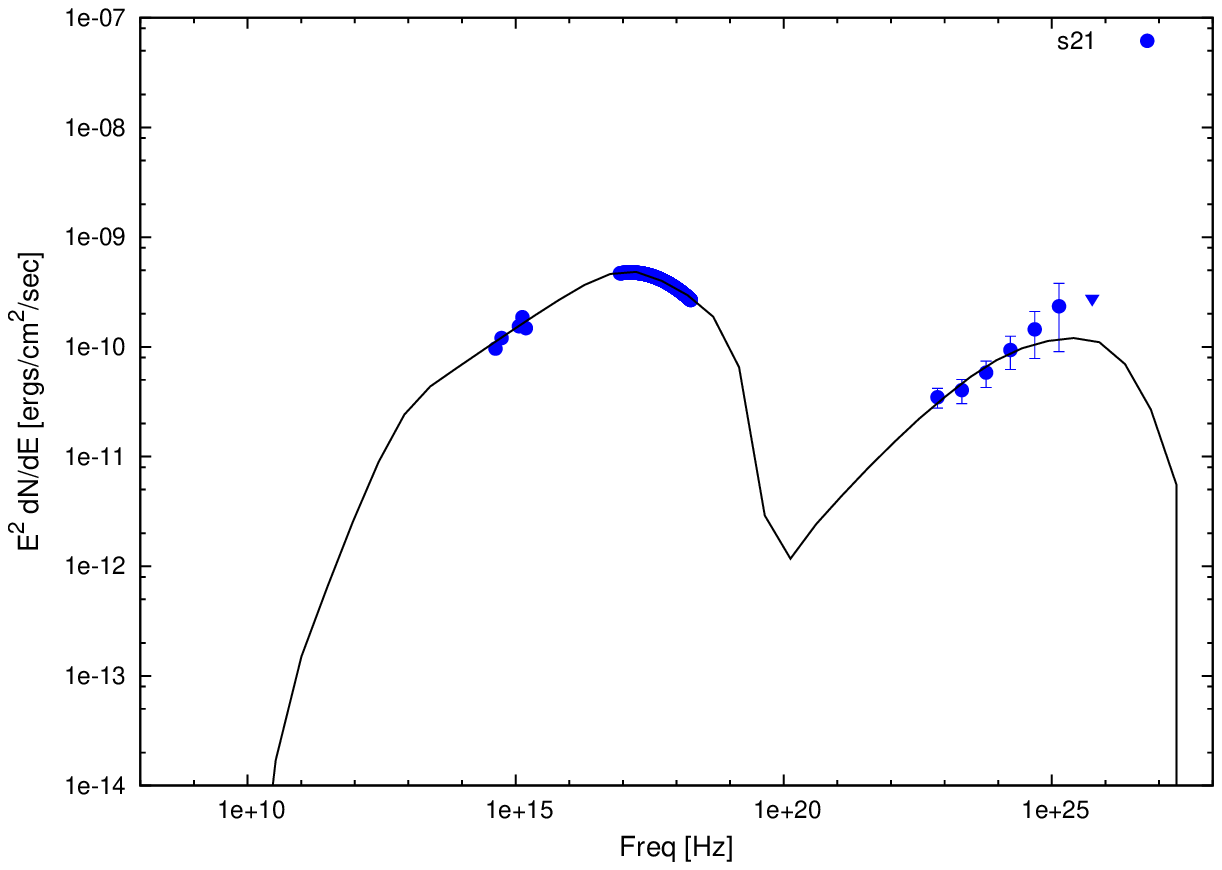}}
      \qquad
  	\subfloat[Model SEDs for all states]{\includegraphics[scale=0.6]{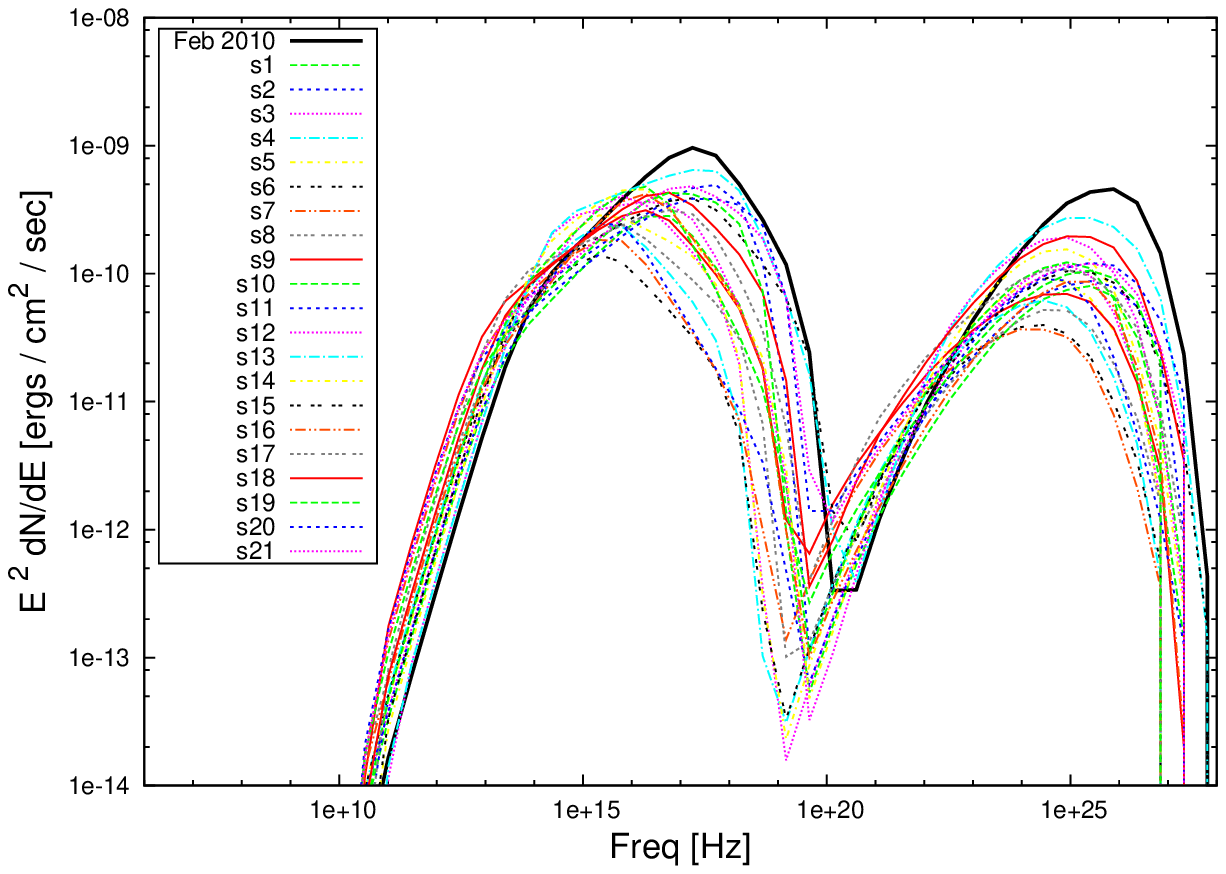}}
	\caption{SED during the different states fit with a one zone SSC. Inverted triangles correspond to upper limits. The last panel shows all the model SEDs for all 21 states along with the model SED during the February 2010 flare (\citet{Amit421}, plotted in thick black line for reference).}
	\label{fig:SED}
\end{figure*}

The ratio of the electric ($U_E$) and the magnetic ($U_B$ )field energy density is parameterised by the equipartition parameter 
$\eta = U_E/U_B$, where $U_B = B^2/2\mu_0$ and $U_E  = m_e c^2 \int\limits_{\gamma_{min}}^{\gamma_{max}} \gamma N(\gamma) d \gamma$. The results of the fit are shown in Fig. \ref{fig:SED}. The best fit parameters and the reduced $\chi^2$ are listed in Table \ref{fitpar}. During all the epochs, the energy density in the jet is matter dominated, with the model parameters a factor of $10$ away from equipartition. However, this ratio may be improved by choosing a lower value of $\delta$ or $R$. For all the SEDs studied here, the one zone model provides a very good fit to the observed data, and does not motivate the need for a second emitting region.

\section[sect:Disc]{Results and Discussions}\label{sect:discussion}

\begin{figure*}
	\centering
	\subfloat[$p_1$ vs $p_2$]{\includegraphics[scale=0.5]{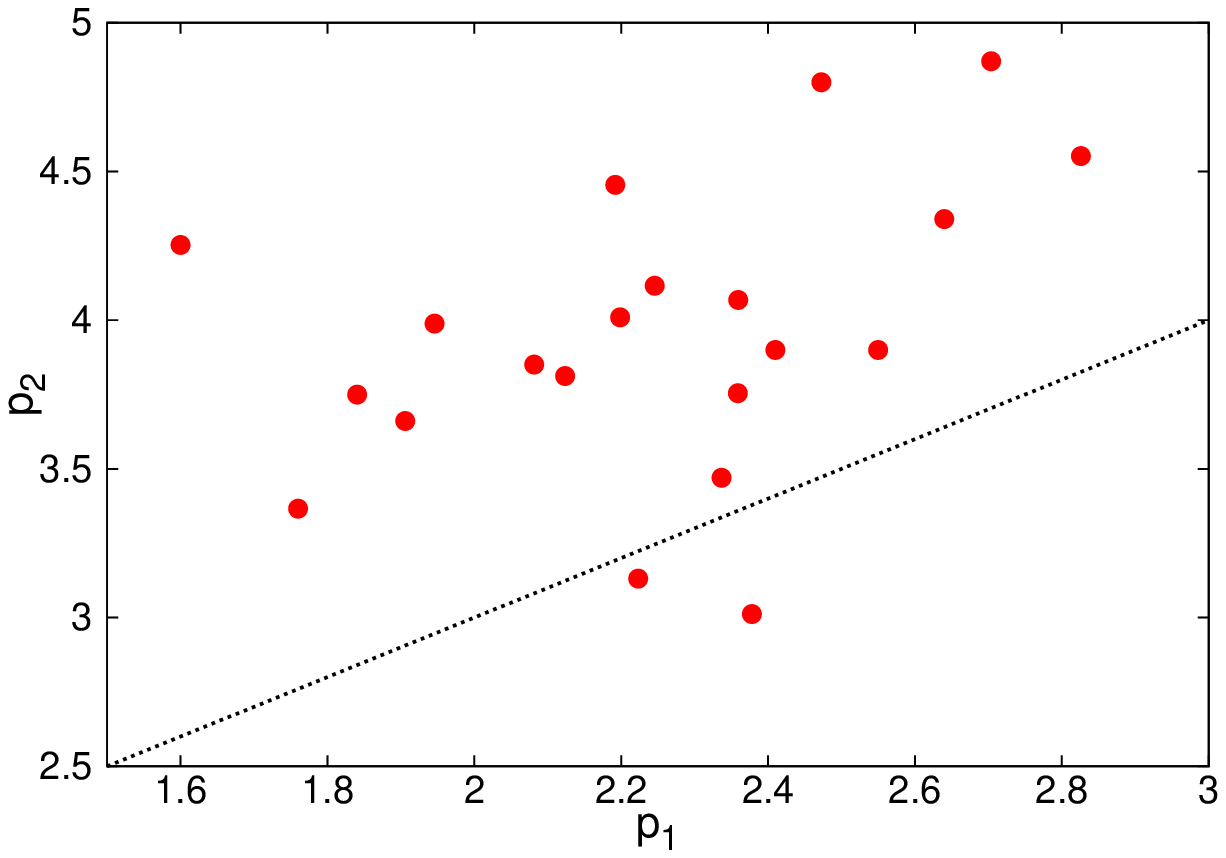}\label{fig:p1p2}}
      \qquad
	  \subfloat[Luminosity vs $\gamma_b$]{\includegraphics[scale=0.5]{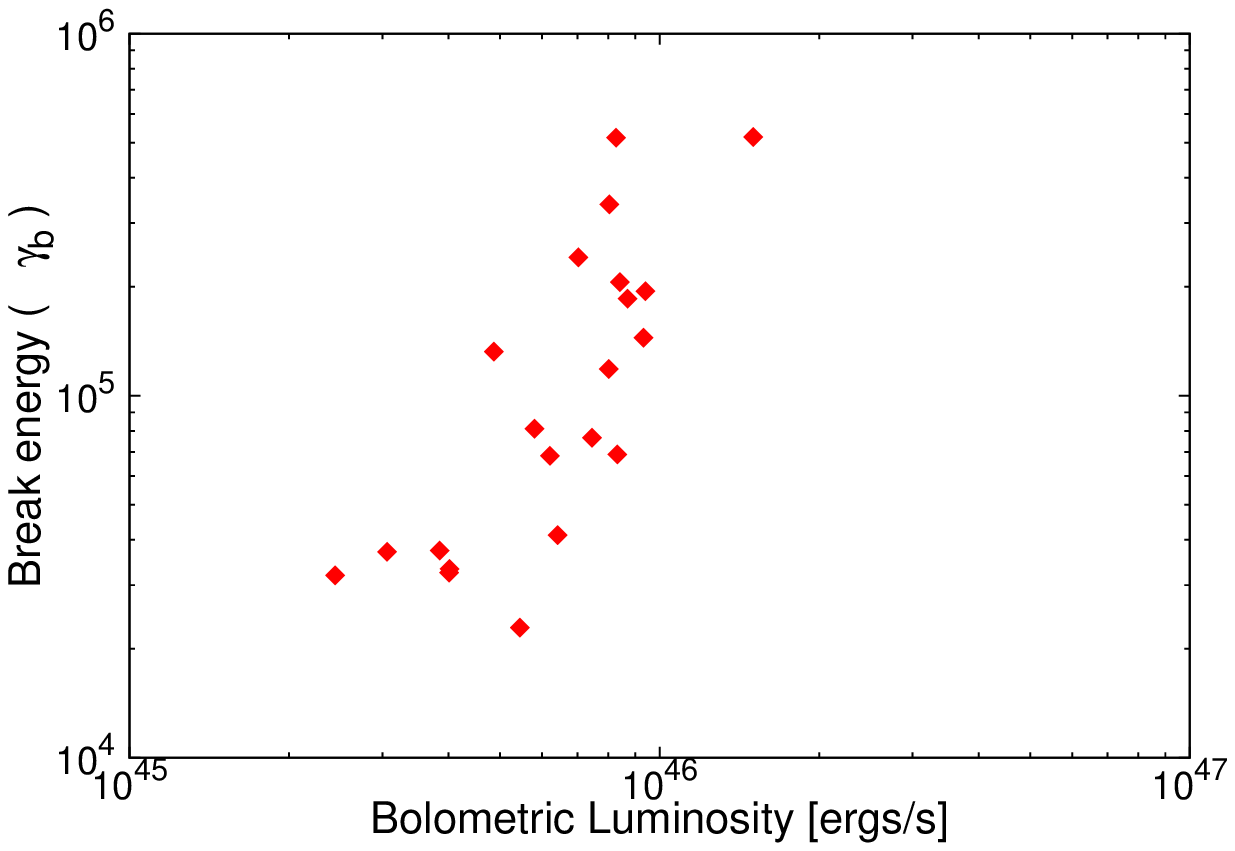}\label{fig:lumgb}}
	  \caption{Cross plot of the fit parameters for the one zone SSC. The black dotted line in the first corresponds to the relation $p_2 = p_1 +1$, which is what would have been expected from a synchrotron cooling break. The second panel shows the variation of the synchrotron peak frequency with the total bolometric luminosity. A strong correlation is clearly observed.}
	\label{fitfig}
\end{figure*}

The similar variability in the Optical-GeV bands, and the X-ray-VHE bands indicates a similar origin for the Optical and Fermi-LAT bands, and for the X-ray-VHE bands. In the framework of the SSC model, this is attributed as the first index, $p_1$ contributing to the  Optical-GeV bands (synchrotron and SSC respectively), and the second index $p_2$ for the X-ray-VHE bands. As can be seen from the fit parameters in Table \ref{fitpar}, some of the states require a very hard particle index $p_1 < 2.0 $. Such an index cannot originate from a single shock acceleration, but may originate from multiple shocks \citep{hard_shock}, which can produce particle indices as hard as 1.5, which is close to the hardest spectrum ($p_1 = 1.6$) obtained in our fits. Alternatively, stochastic acceleration in resonance with plasma wave turbulence behind the relativistic shock front can also harden the injection index \citep{stocast}. The two indices, $p_1$ and $p_2$, show a weak correlation with $\rho = 0.42$, prob = 0.05 (Fig. \ref{fig:p1p2}). In particular, $p_2$ is much steeper than what would be expected from $p_2 = p_1 +1$, thus ruling out the broken power law spectrum as originating from a cooling break. There is a significant correlation between the total bolometric luminosity $L$ and the first index ($\rho = 0.66$, prob = 0.001), and a very strong correlation between $L$ and the particle break energy $\gamma_b$ ($\rho = 0.79$, prob = $3.2 \times 10^{-5}$) (Fig \ref{fig:lumgb}). However, no significant trend is seen between the magnetic field $B$ and $L$. This implies that the variations in the different flux states occur mainly due to a change in the underlying particle distribution, rather than in the jet environment. 

We observe that the $\gamma$-ray flares lead the radio flares, $\Delta t^{obs}_{\gamma,radio} = 52 \pm 20$ days. \citet{Pushkarev} have explained this due to the synchrotron opacity in the nuclear region. Since the radio core is optically thick to synchrotron radiation upto a frequency dependant radius $r_c \propto \nu^{-1}$, the $\gamma$-ray emission zone should be located upstream with respect to the apparent radio core position. Assuming the flares to be caused due the same disturbance, the radio and the $\gamma$-ray flares are not observed simultaneously due to the opacity effect. This information can be used to estimate the distance travelled by emission region $d_{\gamma, radio}$ using the relation \citep{radgamloc}

\begin{equation}
	d_{\gamma,radio} = \frac{\beta_{app} c \Delta t^{obs}_{\gamma,radio}}{sin(\theta)(1+z)}
	\label{eq:dist}
\end{equation}

Using $\beta_{app} \sim 1.92$ \citep{viewang}, $\theta = tan^{-1}(2\beta_{app}/(\beta^2_{app} + \delta^2 -1)) \sim 0.5^\circ $ \citep{ghis_theta_beta} and $z=0.031$, we find the distance travelled by the emission region   $d_{\gamma,radio} = 9.3pc$. This corresponds to a projected distance of $0.08 pc$, equivalent to an angular size of $0.12mas$. \citet{zhang} studied the VLBI image of this source at a resolution of $\sim 0.15 mas$ and found unresolved core. This, compared to our estimated size of $0.12 mas$ allows us to constrain the $\gamma$-ray emission site within the radio core. 

The observed lognormality of the flux distributions in the different bands, and the proportionality between the fluctuations in the flux imply that the variations are lognormal. Similar trends have been found in in several accreting galactic sources \citep{giebels_lognorm} and also in the Seyfert 1 galaxy, Mkn 766, where the physical process responsible for the X-ray emission is most likely thermal emission from the galactic disk \citep{Vaughan}. The results of our SED modeling in Section \ref{sect:spectra} show that the flux variability in the source can be mainly attributed to changes in the particle spectrum, and not the jet parameters like the magnetic field or the Doppler factor. This implies that the lognormal fluctuations in the flux are indicative of lognormal processes in the accretion disk, and the lognormal fluctuations in the accreting rate give rise to an injection rate with similar properties.

The one zone SSC model used in this study provides a suitable fit to the observed SEDs for all epochs, without motivating the need for any additional emitting regions. Interestingly, while the time-resolved UV-X-ray spectra during the April 2013 flare could not be explained by synchrotron emission from a single region \citep{my421}, the time averaged broadband spectrum during the same time (State s13) can be well fitted by our model (Reduced $\chi^2 = 1.5$). This motivates a need for better time resolved broadband spectra. The upcoming Major Atmospheric Cherenkov Experiment (MACE) \citep{MACE} at Hanle, Ladakh is expected to provide us with excellent time resolved spectrum at VHE energies, which may throw new light on the VHE emission from blazar jets. 

\begin{acknowledgements}
We thank the scientific and technical personnel of HAGAR groups at IIA, Bengaluru and TIFR, Mumbai for observations with HAGAR system.
A. Sinha would like to thank Sunder Sahayanathan from the Bhabha Atomic Research Center, Mumbai, for helpful discussions and comments. This research has made use of data,
software and/or web tools obtained from NASAs High Energy Astrophysics Science Archive
Research Center (HEASARC), a service of Goddard Space Flight Center and the Smithsonian
Astrophysical Observatory. Part of this work is based on archival data, software, or online
services provided by the ASI Science Data Center (ASDC).        This research has made use of the
  XRT Data Analysis Software (XRTDAS) developed under the responsibility
the ASI Science Data Center (ASDC), Italy, and the  NuSTAR Data Analysis Software (NuSTARDAS) jointly developed by the ASI Science
    Data Center (ASDC, Italy) and the California Institute of Technology (Caltech, USA). 
        Data from the Steward Observatory spectropolarimetric monitoring project are used. This
program is supported by Fermi Guest Investigator grants NNX08AW56G, NNX09AU10G, and
NNX12AO93G. The OVRO 40 M Telescope Fermi Blazar Monitoring Program is supported by NASA under awards 
NNX08AW31G and NNX11A043G, and by the NSF under awards AST-0808050 and AST-1109911.

\end{acknowledgements}

\bibliography{Mkn421,hagar}
\bibliographystyle{aa}

\begin{table*}
		\caption{The 21 states contemporaneous with HAGAR time periods for which SED has been constructed. The Crab flux corresponds to 3.9 counts/min.}
        \begin{tabular}{c c c c c c c c c}
                \hline
                \hline
				state &  \multicolumn{2}{c}{Start date} & \multicolumn{2}{c}{End date}    &   Total Duration   & Excess counts  &  Count rate  &  Significance  \\ 
							   &  ISO  &  MJD & ISO & MJD &  mins & & counts/min & \\ \hline  
s1   & 2009-03-21 & 54911 & 2009-05-01 & 54952 &   480  &  2095.3  $\pm$ 517.0     &    4.36   $\pm$   1.07 & 4.05 \\  
s2   & 2010-03-11 & 55266 & 2010-03-18 & 55273 &   600  &  2938.2  $\pm$ 579.1     &    4.91   $\pm$   0.97 & 5.1    \\  
s3   & 2011-01-02 & 55563 & 2011-01-12 & 55573 &   356  &  1304.1  $\pm$ 455.6     &    3.65   $\pm$   1.27 & 2.86  \\  
s4   & 2011-01-29 & 55590 & 2011-02-10 & 55602 &   558  &  2466.2  $\pm$ 563.2     &    4.40   $\pm$   1.10 & 4.37  \\  
s5   & 2011-03-01 & 55621  & 2011-03-10 & 55630 &   365  &  2425.7  $\pm$ 451.6     &    6.63   $\pm$   1.23 & 5.36  \\  
s6   & 2012-01-19 & 55945	& 2012-01-31 & 55957  &   401  &  40.1    $\pm$ 498.1     &    0.10   $\pm$ 1.24  &   0.08   \\  
s7   & 2012-02-16 & 55973	& 2012-02-28 & 55985 &   160  & -241.9   $\pm$ 293.3     &   -1.50   $\pm$ 1.82  &   0.82   \\  
s8   & 2012-03-15 & 56001	& 2012-03-23 & 56009 &   197  &  61.2    $\pm$ 332.3     &    0.30   $\pm$ 1.67  &   0.18   \\  
s9   & 2012-04-12 & 56029	& 2012-04-18 & 56035 &   113  & -141.3   $\pm$ 256.7     &   -1.24   $\pm$ 2.25  &   0.55   \\  
s10  & 2013-01-09 & 56301	& 2013-01-16 & 56308 &   204  & -389.6  $\pm$ 390.4     &   -1.91   $\pm$ 1.90  &   1.0   \\  
s11  & 2013-02-08 &	56331 & 2013-02-17 & 56340 &    241 &  416.0   $\pm$ 436.2     &    1.72   $\pm$ 1.80  &   0.95   \\  
s12  & 2013-03-04 & 56355 & 2013-03-15 & 56366 &    326 &  1377.1  $\pm$ 442.6     &    4.21   $\pm$ 1.35  &   3.11   \\ 
s13  & 2013-04-01 & 56383 & 2013-04-15 & 56397 &    120 &  825.6   $\pm$ 277.7     &    6.85   $\pm$ 2.30  &     3.00   \\
s14  & 2013-05-01 & 56413	& 2013-05-13 & 56425 &    279 &  436.8   $\pm$ 452.3     &    1.56   $\pm$ 1.61  &   0.96   \\ 
s15  & 2013-12-29 & 56655	& 2014-01-07 & 56664 & 	316 & -290.7   $\pm$ 469.9     &   -0.91   $\pm$ 1.48  &   0.61   \\  
s16  & 2014-01-28 & 56685	& 2014-02-07 & 56695 &    180 & -444.2   $\pm$ 300.6     &   -2.46   $\pm$ 1.66  &   1.47   \\  
s17  & 2014-04-01 & 56748	& 2014-04-14 & 56761 &    306 &  837.1   $\pm$ 416.6     &    2.72   $\pm$ 1.35  &   2.00   \\  
s18  & 2014-04-21 & 56768	& 2014-04-29 & 56776 &    60  &  423.3   $\pm$ 188.8     &    7.17   $\pm$ 3.19  &   2.24   \\ 
s19  & 2015-01-18 & 57040 & 2015-01-28 & 57050 &    184 &  2003.9  $\pm$ 377.1     &   10.84   $\pm$ 2.04  &   5.31   \\  
s20  & 2015-02-12 & 57065 & 2015-02-23 & 57076 &    301 &  2690.9  $\pm$ 454.5     &    8.93   $\pm$ 1.50  &   5.91   \\  
s21  & 2015-04-12 & 57124 & 2015-04-23 & 57135 &    301 &  -49.4   $\pm$ 438.5     &   -0.16   $\pm$ 1.45  &   0.11   \\  				
                \hline               
        \end{tabular}
        \label{hagarcounts}
\end{table*}

\begin{table*}
	\caption{Fractional variability ($F_{var}$) in different wavebands with different time binnings.}
	\begin{tabular}{l c c c c}
		\hline
		&    \multicolumn{4}{c}{Time Binning} \\
		Waveband     			& 1 day & 10 days & 1 month & 3 months \\
		\hline
		OVRO (15GHz) 	  		& $ 0.22 \pm 0.01  $ &  $ 0.20 \pm 0.01 $ & $ 0.19 \pm 0.01  $ & $ 0.19 \pm 0.01 $  \\    
		Optical R    			& $ 0.32 \pm 0.01  $ &   $ 0.31 \pm 0.01 $ & $ 0.31 \pm 0.01  $ & $ 0.29 \pm 0.02 $  \\
		Optical V    			& $ 0.37 \pm 0.01  $ &   $ 0.36 \pm 0.01 $ & $ 0.37 \pm 0.01  $ & $ 0.35 \pm 0.02 $  \\
		UV UW1 					& $ 0.44 \pm 0.01  $ &   $ 0.39 \pm 0.02 $ & $ 0.37 \pm 0.01  $ & $ 0.37 \pm 0.03 $  \\
		UV UM2 		 			& $ 0.47 \pm 0.01  $ &   $ 0.42 \pm 0.02 $ & $ 0.41 \pm 0.02  $ & $ 0.40 \pm 0.03 $  \\
		UV UW2 		 	     	& $ 0.48 \pm 0.01  $ &   $ 0.43 \pm 0.03 $ & $ 0.40 \pm 0.03  $ & $ 0.41 \pm 0.03 $ \\
		Swift-XRT (0.2-10keV)	& $ 0.71 \pm 0.01 $ &   $ 0.61 \pm 0.04 $ & $ 0.57 \pm 0.04  $ & $ 0.54 \pm 0.03 $ \\
		MAXI (2-20keV)			& $ 0.81 \pm 0.01 $ &   $ 0.79 \pm 0.10 $ & $ 0.74 \pm 0.08  $ & $ 0.67 \pm 0.02 $  \\
		Swift-BAT (15-150keV)	& $ 1.08 \pm 0.01 $ &   $ 0.95 \pm 0.16 $ & $ 0.90 \pm 0.10  $ & $ 0.81 \pm 0.04 $  \\
        Fermi-LAT (0.2-2GeV)  	& $ - $ &   $ 0.44 \pm 0.08 $ & $ 0.41 \pm 0.06  $ & $ 0.38 \pm 0.05  $  \\
        Fermi-LAT (2-300GeV)  	& $ - $ &   $ 0.48 \pm 0.14$ & $ 0.42 \pm 0.08   $ & $ 0.38 \pm 0.07 $  \\
		HAGAR ( $>200$GeV)      & $ - $ &   $ 1.07 \pm 0.24$ & $1.07 \pm 0.24    $ & $ - $  \\
		\hline 
	\end{tabular}
	\label{tab:fvar}
\end{table*}

\begin{table*}
	\caption{Reduced $\chi^2$ for the fit of a Gaussian function to the flux distribution (column 2); Lognormal distribution to the flux distribution (coulmn 3); 
		A constant line to the excess variance vs. flux (column 4) and a linear function (column 5)  for the excess variance vs. flux in the various wavebands. 
	Columns 6 and 7 give the Pearsons $r$ and Spearmans $\rho$ correlation coefficients respectively.
	The values in the brackets denote the null hypothesis probabilities.}
	\begin{tabular}{c c c c c c c }
		\hline
		\hline
		& Gaussian & Lognormal & Constant  &  Linear  &  $r$ (prob) & $\rho$ (prob) \\
		\hline
		OVRO (15 GHz)			 & 45.3 & 36.2  &    3.4  & 2.8   &  0.48 (0.12)   & 0.58 (0.01)      \\ 
		SPOL (R band)   		 & 14.9 & 10.0  &    4.9  & 4.3   &  0.26 (0.39)   & 0.21 (0.49)      \\ 
		SPOL (V band)   		 & 6.3 & 4.0  &    5.7  & 4.3   &  0.61 (2.9E-3) & 0.51 (0.02)     \\  
		Swift-UW1 				 & 8.3 & 6.7  &    11.3 & 10.1  &  0.24 (0.31)   & 0.14 (0.51)     \\ 
		Swift-UW2 				 & 6.0 & 5.3  &    12.9 & 12.1  &  0.35 (0.14)   & 0.23 (0.33)     \\  
		Swift-UM2 		         & 6.1 & 4.9  &    9.9  & 8.1   &  0.20 (0.39)   & 0.21 (0.21)     \\  
		Swift-XRT (0.2 - 10 keV) & 2.6 & 1.7  &    66.9 & 46.4  &  0.77 (3.7E-4) & 0.72 (1.1e-3)   \\  
		MAXI 	  (2 - 20 keV)   & 19.3 & 1.5  &    2.2  & 0.95  &  0.86 (2E-16)  & 0.91 (7.9E-8)  \\  
		Swift-BAT (15-150 keV)   & 3.8	& 0.9	& 	   1.7  & 0.82  &  0.78 (9.0E-4) & 0.90 (4.4E-6)  \\    
		Fermi-LAT (0.2-300 GeV)  & 1.9	& 1.2	& 	   2.5  & 1.6   &  0.57 (4.8E-3) & 0.74 (4.5E-5)  \\ 
		\hline
	\end{tabular}
	\label{lognorm}
\end{table*}

\begin{table*}
	\caption{Fit parameters for the different SED states, with $R=6.0 \times 10^{16}$ cm, $\delta = 21$, $\gamma_{min}=1000$ and  $\gamma_{max}=10\gamma_b$}
	\begin{tabular}{ccccccccc}
		\hline
		\hline
		& $p_1$ & $p_2$ & ln($\gamma_b$) & $B$(G) & $U_E$(ergs/cc) & $\eta = U_E/U_B$ & $L$ (ergs/sec) & $\chi^2$ \\ 
		\hline
s1  &  2.2  & 4.0  & 11.79  & 2.94E-02  & 4.53E-04  & 13.1  & 4.86E+45  & 1.65 \\
s2  &  2.4  & 4.8  & 13.15  & 3.79E-02  & 4.47E-04  & 7.79  & 8.28E+45  & 0.98 \\
s3  &  1.8  & 3.7  & 10.62  & 4.22E-02  & 4.33E-04  & 6.12  & 6.41E+45  & 0.48 \\
s4  &  2.0  & 3.8  & 10.38  & 3.99E-02  & 4.50E-04  & 7.10  & 4.00E+45  & 0.46 \\
s5  &  1.7  & 3.3  & 10.03  & 3.97E-02  & 4.84E-04  & 7.70  & 5.45E+45  & 0.51 \\
s6 	&  2.1 	& 3.8  & 10.37 	& 2.91E-02 	& 5.08E-04 	& 15.0 	& 2.44E+45 	& 1.26 \\
s7 	&  2.2 	& 4.1  & 10.52 	& 3.99E-02 	& 4.85E-04 	& 7.65 	& 3.06E+45 	& 0.90 \\
s8 	&  1.9 	& 3.6  & 10.41 	& 4.13E-02 	& 3.82E-04 	& 5.63 	& 4.01E+45 	& 1.38 \\
s9 	&  2.4 	& 3.8  & 11.30 	& 4.42E-02 	& 5.56E-04 	& 7.16 	& 5.80E+45 	& 0.90 \\
s10	&  2.1 	& 4.4  & 11.24 	& 4.59E-02 	& 4.48E-04 	& 5.34 	& 7.45E+45  & 1.30 \\
s11 &  1.6  & 4.2  & 10.52  & 3.05E-02  & 5.06E-04  & 13.6  & 3.84E+45  & 2.35 \\
s12 &  2.8  & 4.5  & 12.13  & 3.91E-02  & 6.29E-04  & 10.3  & 8.69E+45  & 1.90 \\
s13 &  2.7  & 4.8  & 13.15  & 3.85E-02  & 6.35E-04  & 10.7  & 1.50E+46  & 1.52 \\
s14	&  2.3 	& 4.0  & 11.14 	& 4.42E-02 	& 4.90E-04 	& 6.29 	& 8.32E+45 	& 1.43 \\
s15	&  2.5 	& 3.9  & 12.73 	& 3.65E-02 	& 4.99E-04 	& 9.40 	& 8.03E+45 	& 1.32 \\
s16	&  1.9 	& 3.9  & 11.13 	& 4.38E-02 	& 3.65E-04 	& 4.77 	& 6.20E+45 	& 1.34 \\
s17	&  2.6 	& 4.3  & 12.39 	& 3.57E-02 	& 7.05E-04 	& 13.8 	& 7.02E+45 	& 0.97 \\
s18	&  2.3 	& 3.7  & 11.88 	& 2.99E-02 	& 8.22E-04 	& 23.1 	& 9.32E+45 	& 1.01 \\
s19 &  2.2  & 3.1  & 11.68  & 3.53E-02  & 5.31E-04  & 10.7  & 8.01E+45  & 1.98 \\
s20 &  2.3  & 3.0  & 12.23  & 2.71E-02  & 8.06E-04  & 27.5  & 8.42E+45  & 1.10 \\
s21 &  2.3  & 3.4  & 12.17  & 3.58E-02  & 6.28E-04  & 12.3  & 9.40E+45  & 1.59 \\
		\hline
	\end{tabular}
	\label{fitpar}
\end{table*}


\begin{figure*}
	\centering
  	\subfloat[s1]{\includegraphics[scale=0.6]{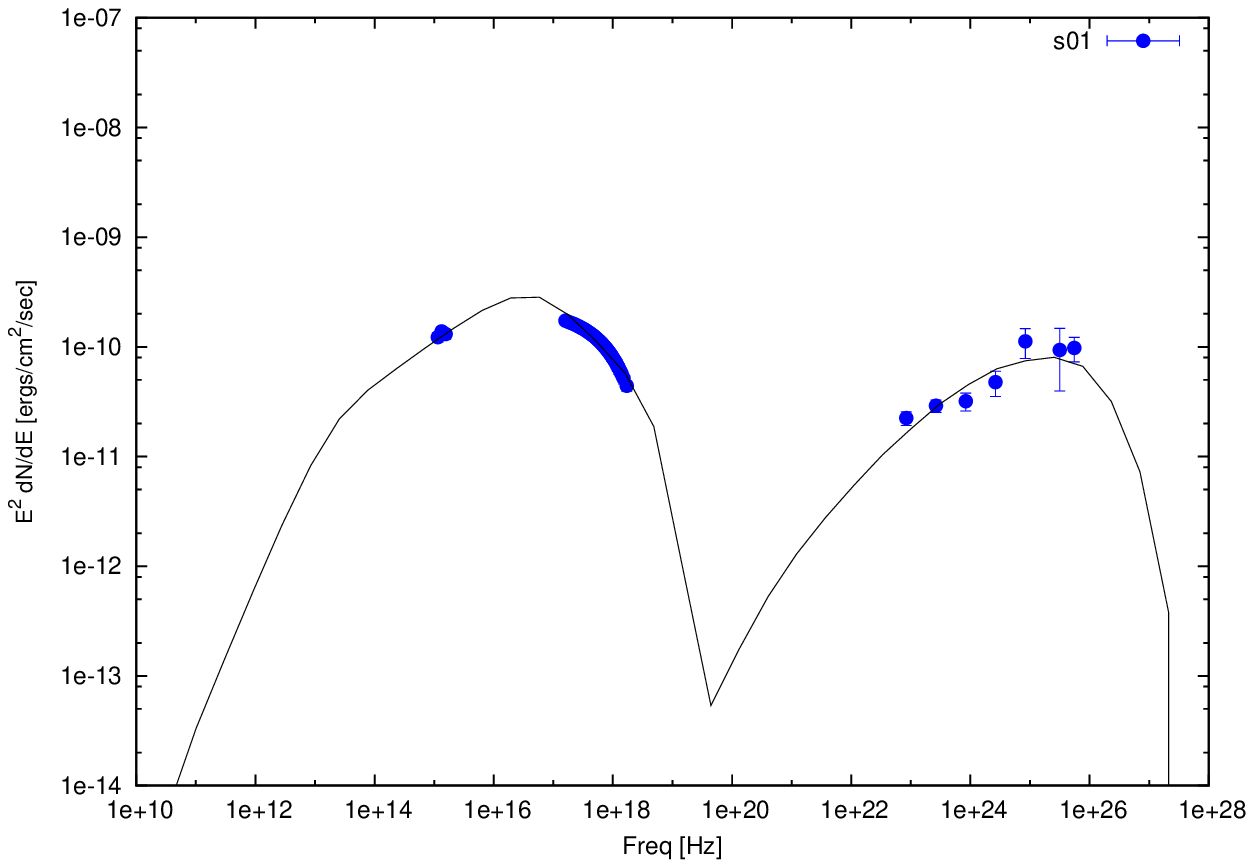}}
      \qquad
 	\subfloat[s2]{\includegraphics[scale=0.6]{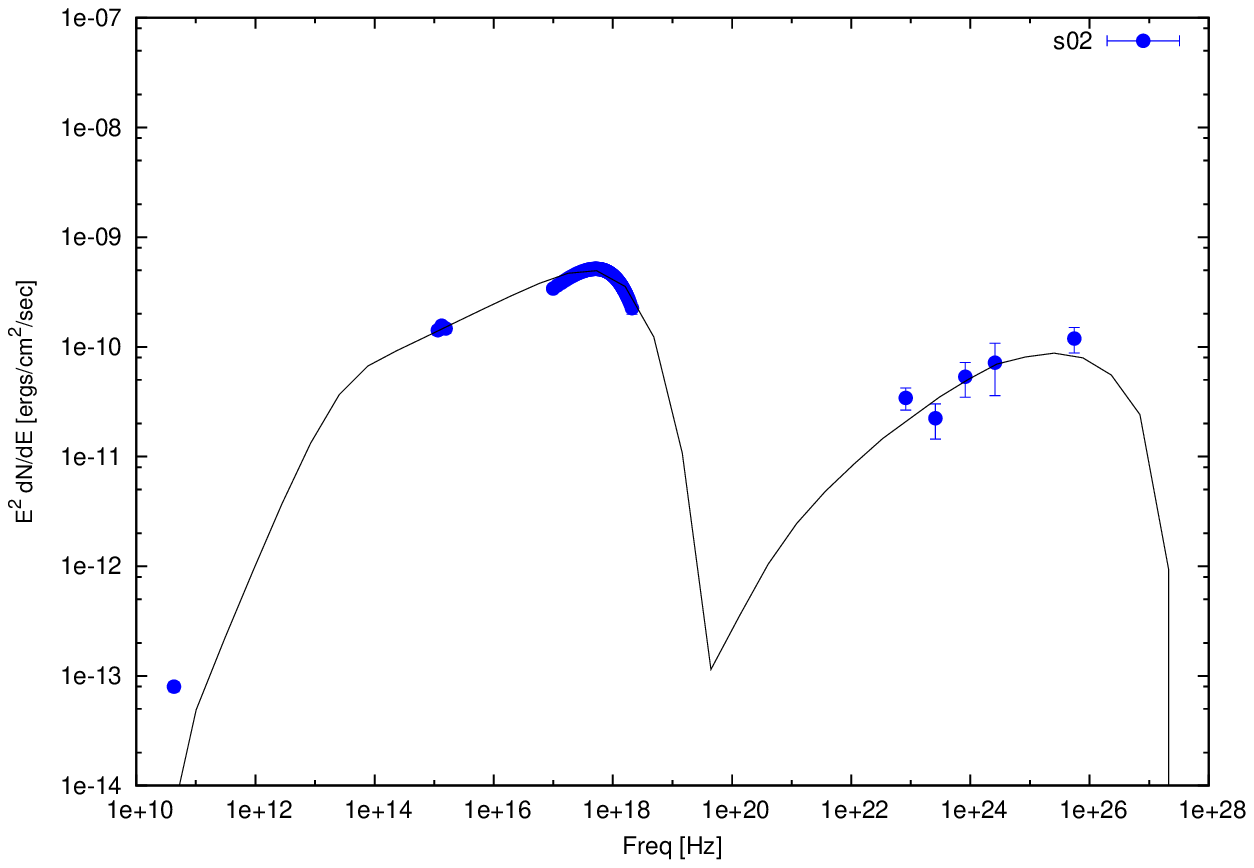}}
      \qquad
  	\subfloat[s3]{\includegraphics[scale=0.6]{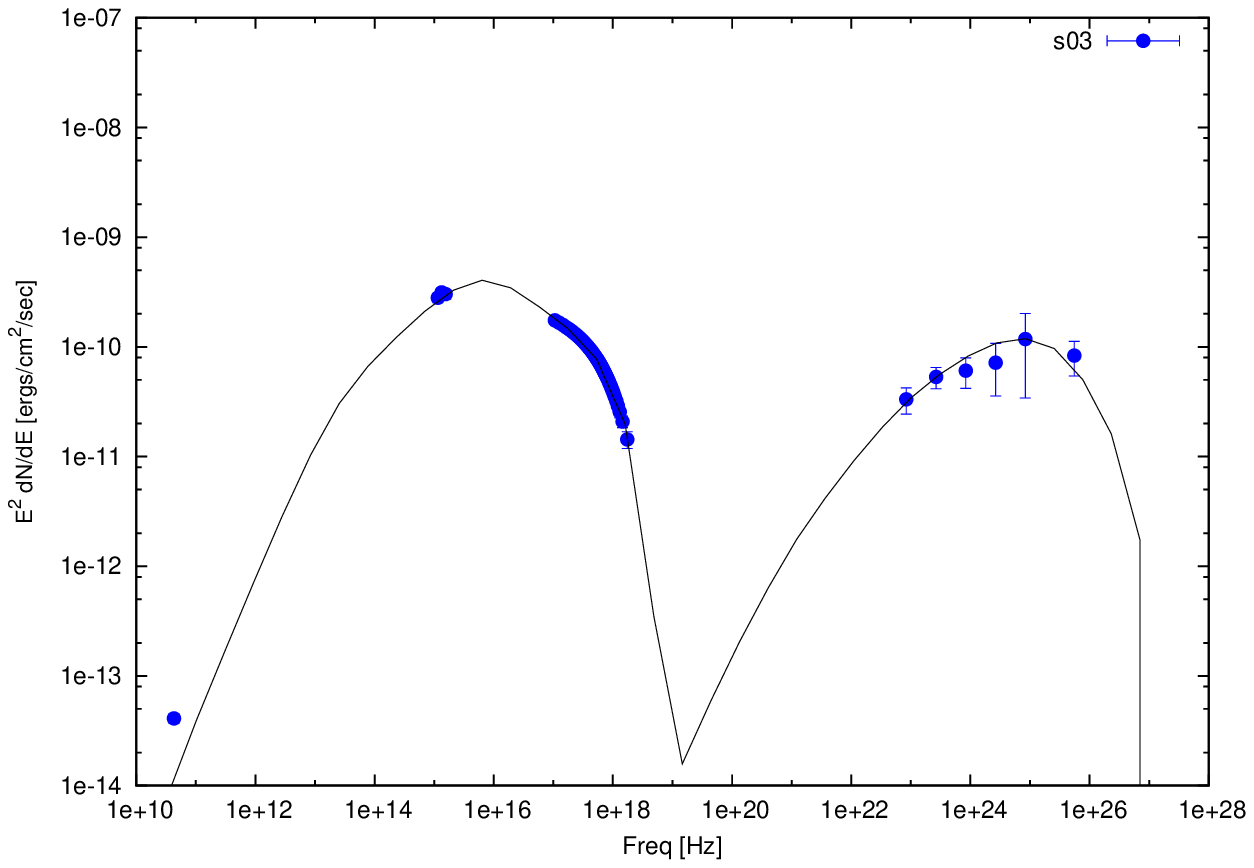}}
      \qquad
  	\subfloat[s4]{\includegraphics[scale=0.6]{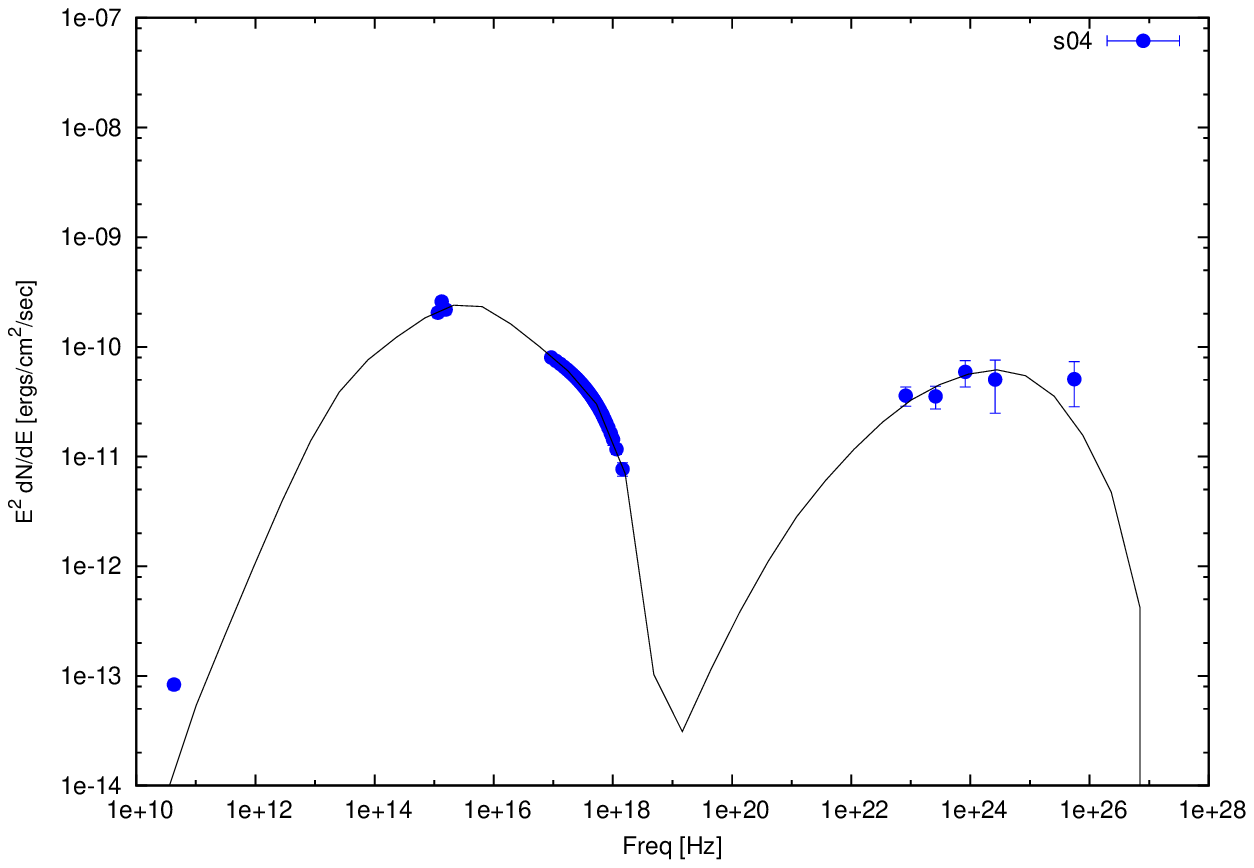}}
      \qquad
  	\subfloat[s5]{\includegraphics[scale=0.6]{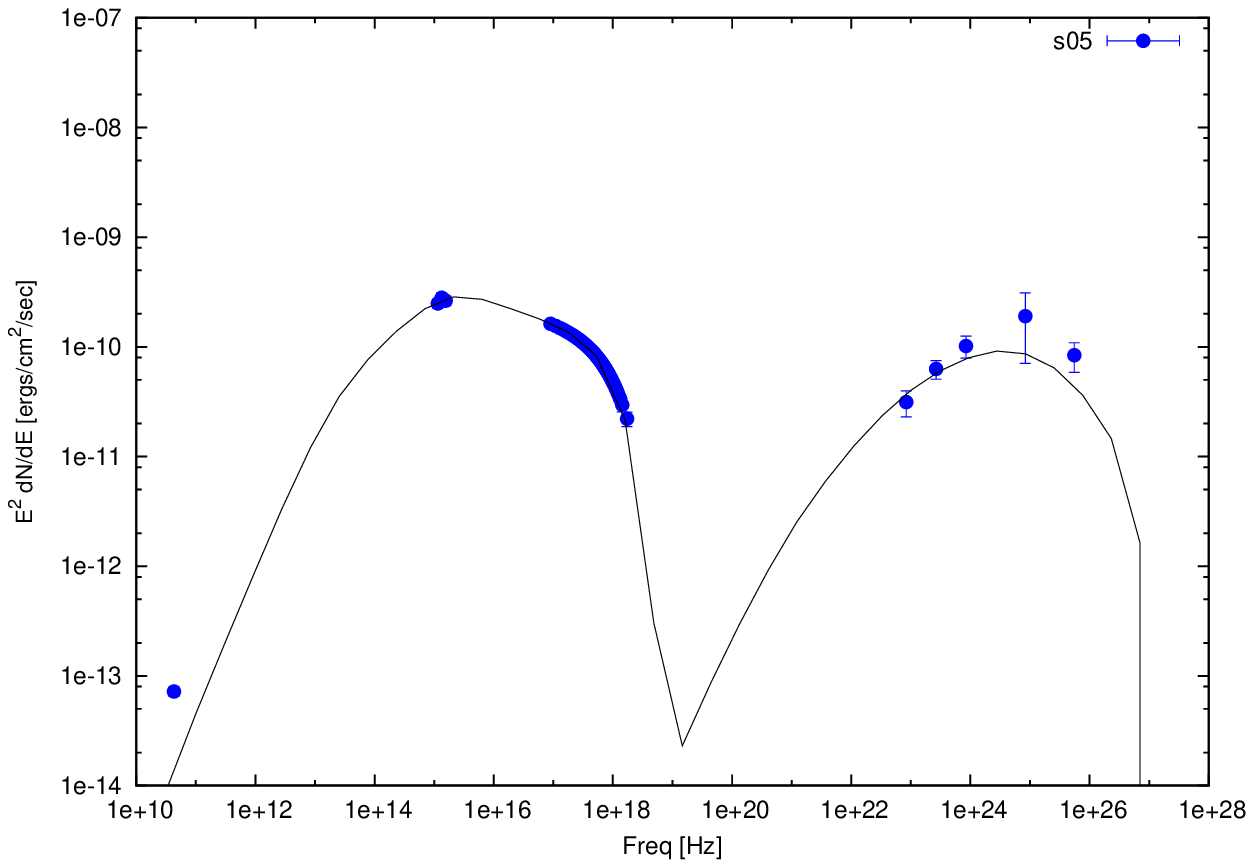}}
      \qquad
  	\subfloat[s6]{\includegraphics[scale=0.6]{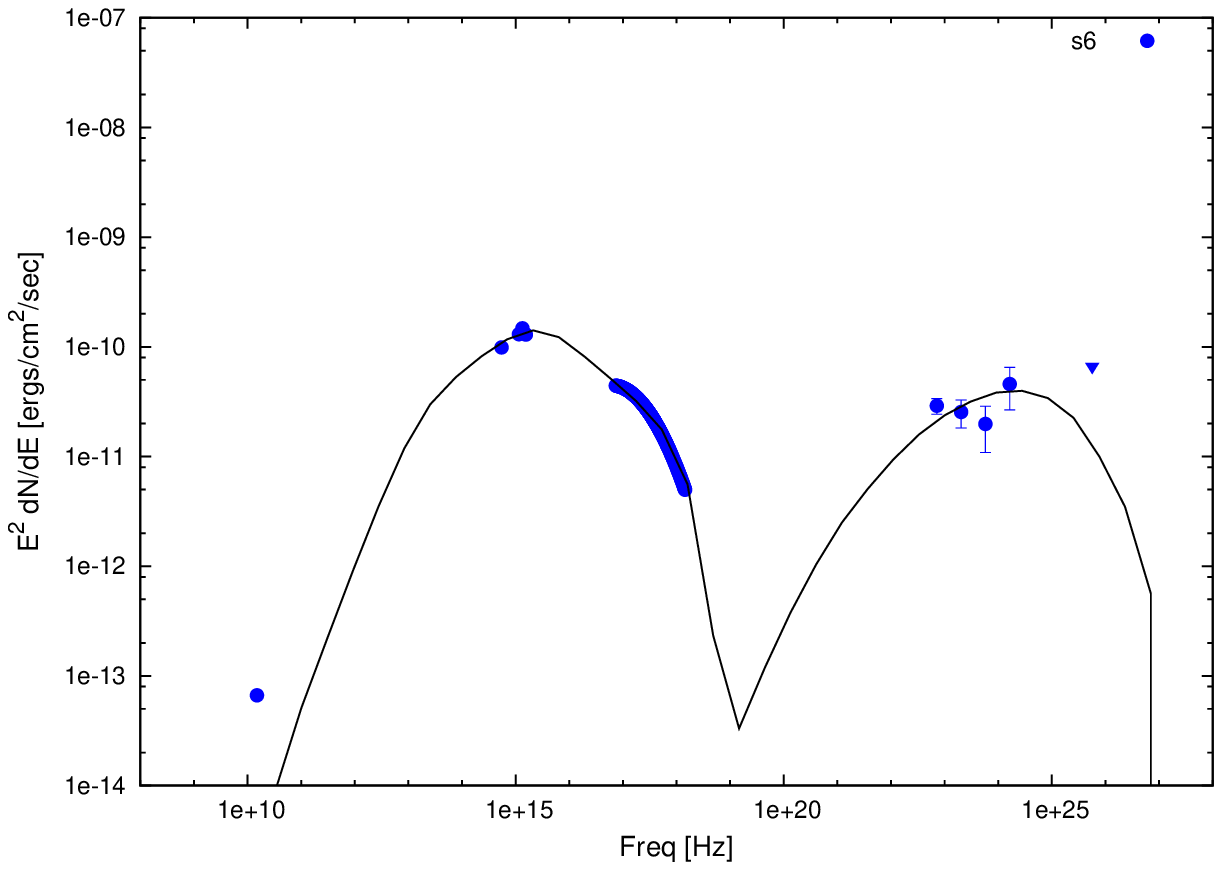}}
      \qquad
  	\subfloat[s7]{\includegraphics[scale=0.6]{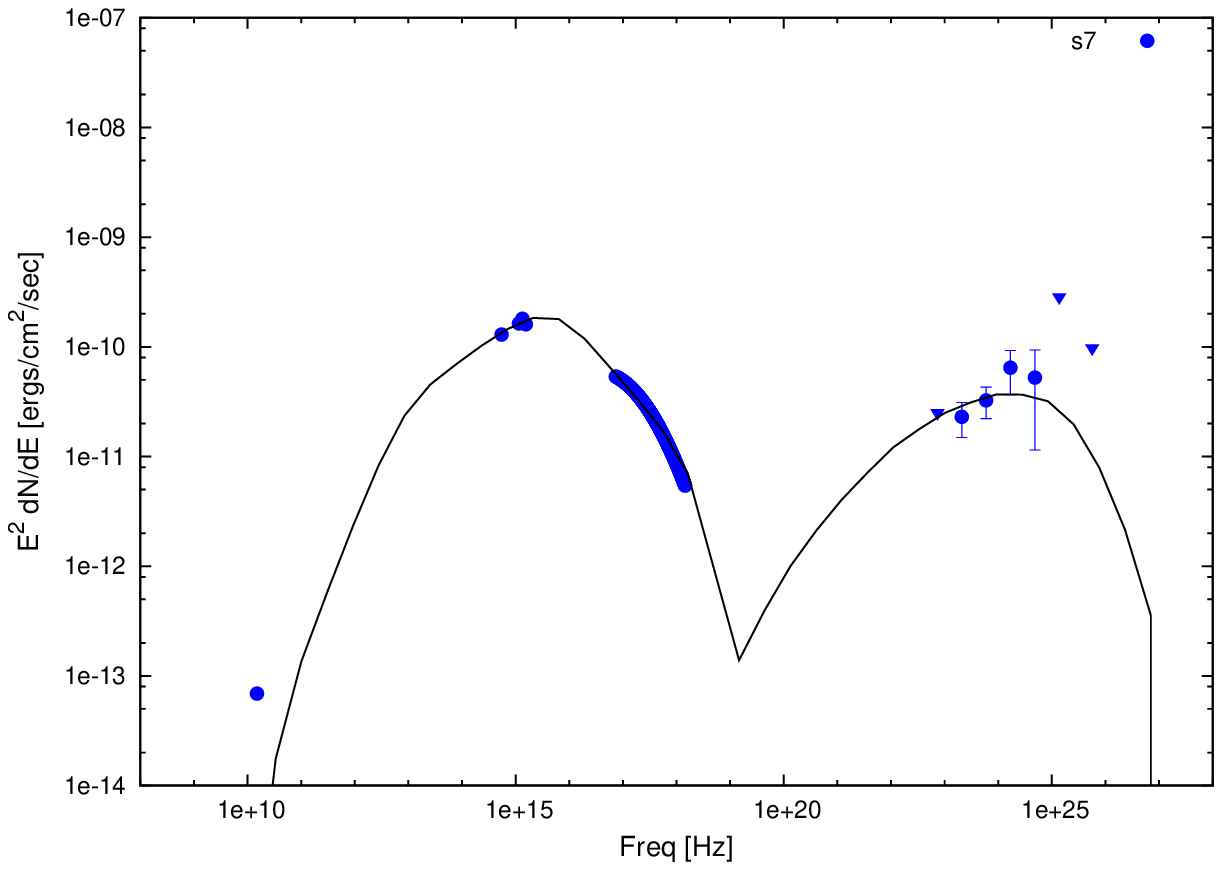}}
      \qquad
  	\subfloat[s8]{\includegraphics[scale=0.6]{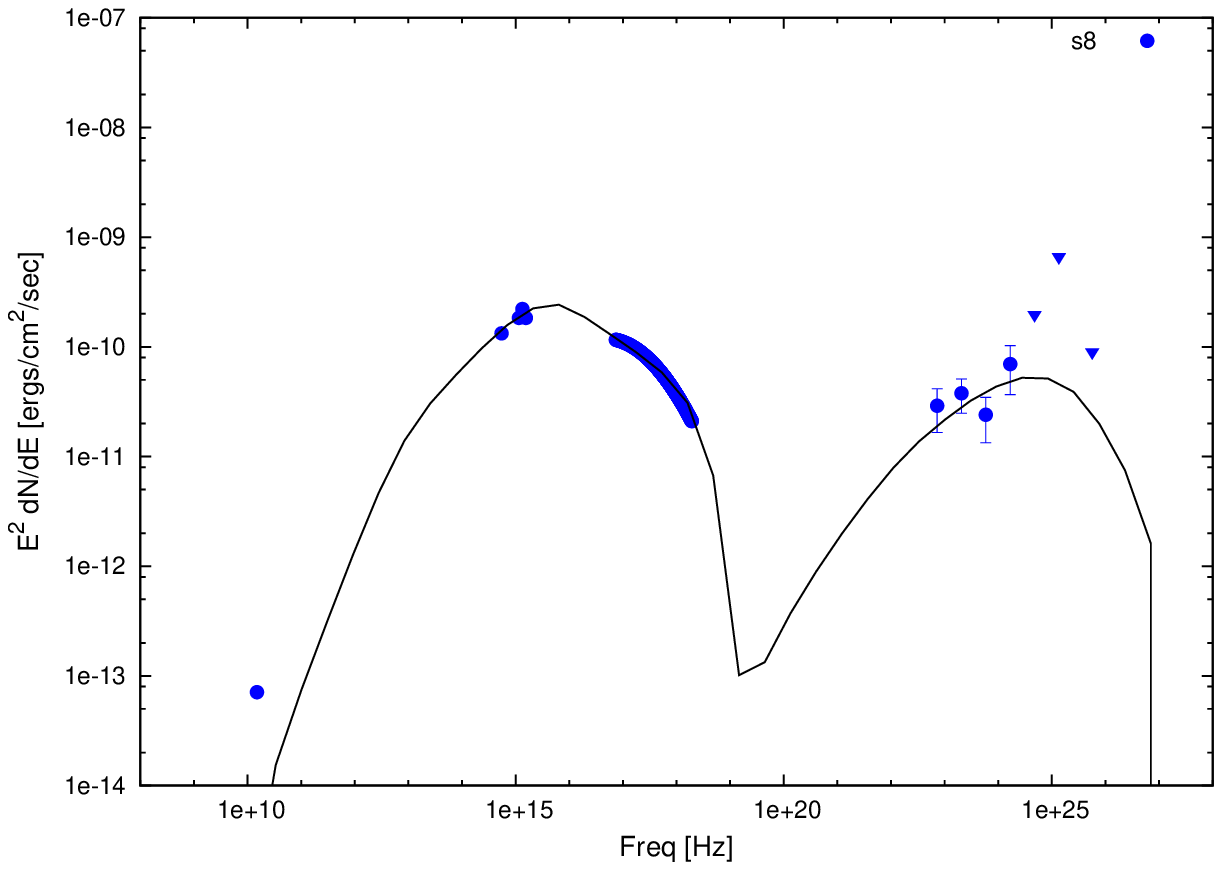}}
      \qquad
  \end{figure*}
  \begin{figure*}
	  \subfloat[s9]{\includegraphics[scale=0.6]{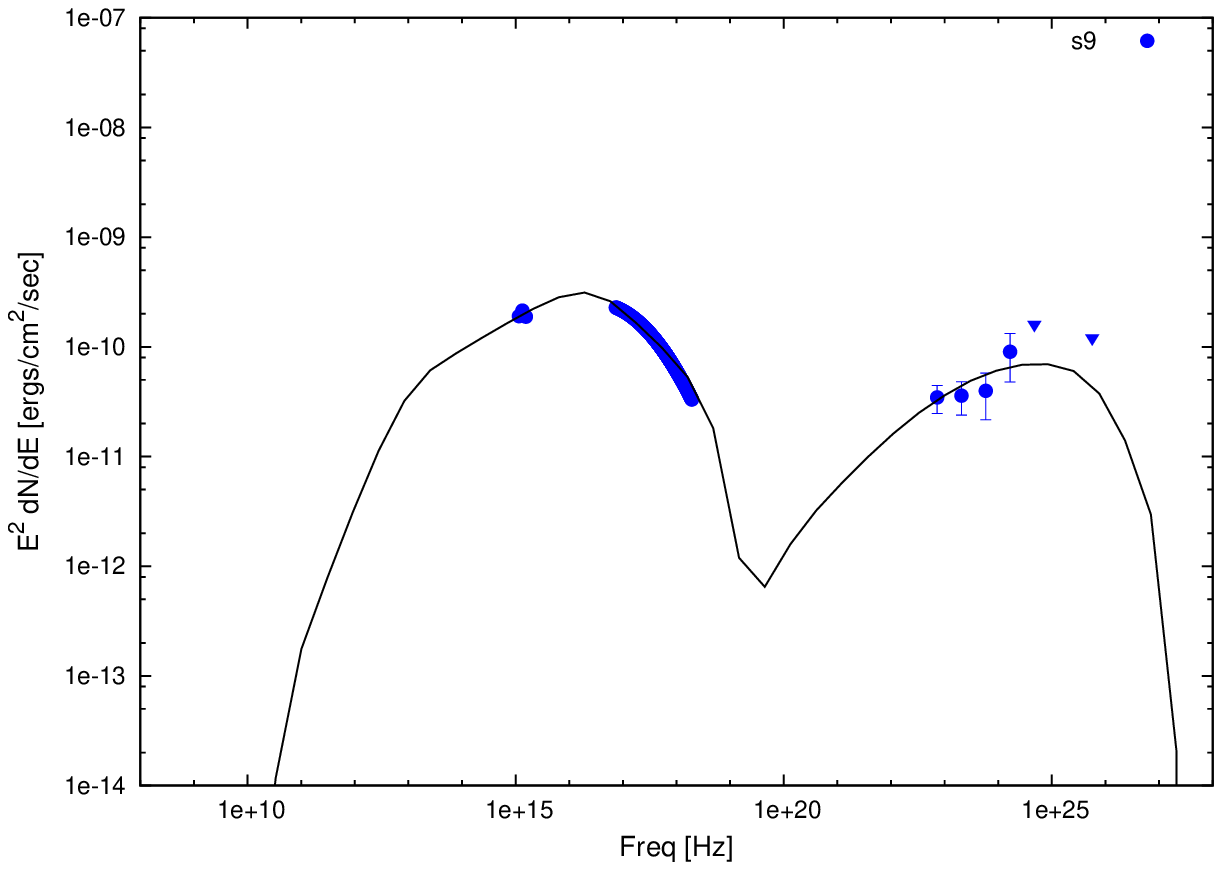}}
      \qquad
	  \subfloat[s10]{\includegraphics[scale=0.6]{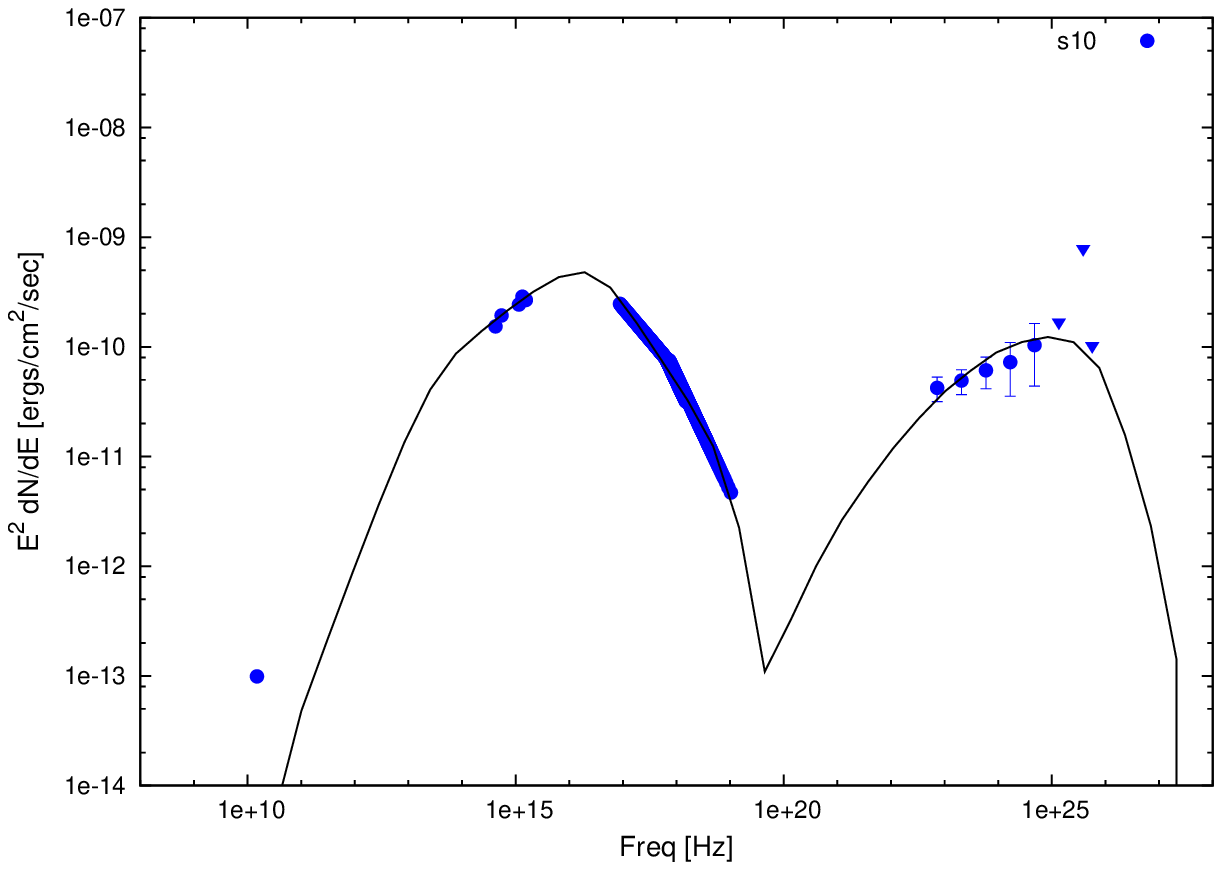}}
      \qquad
  	\subfloat[s11]{\includegraphics[scale=0.6]{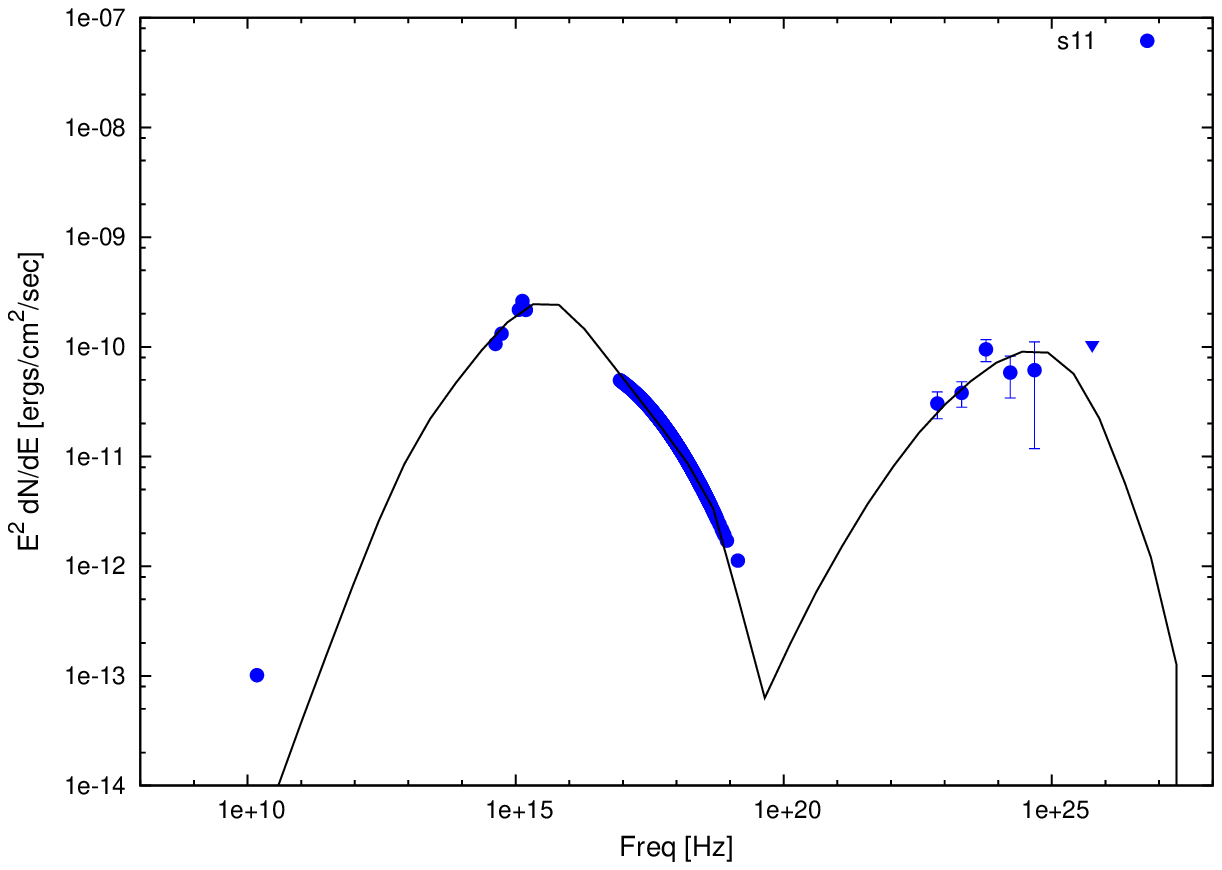}}
      \qquad
  	\subfloat[s12]{\includegraphics[scale=0.6]{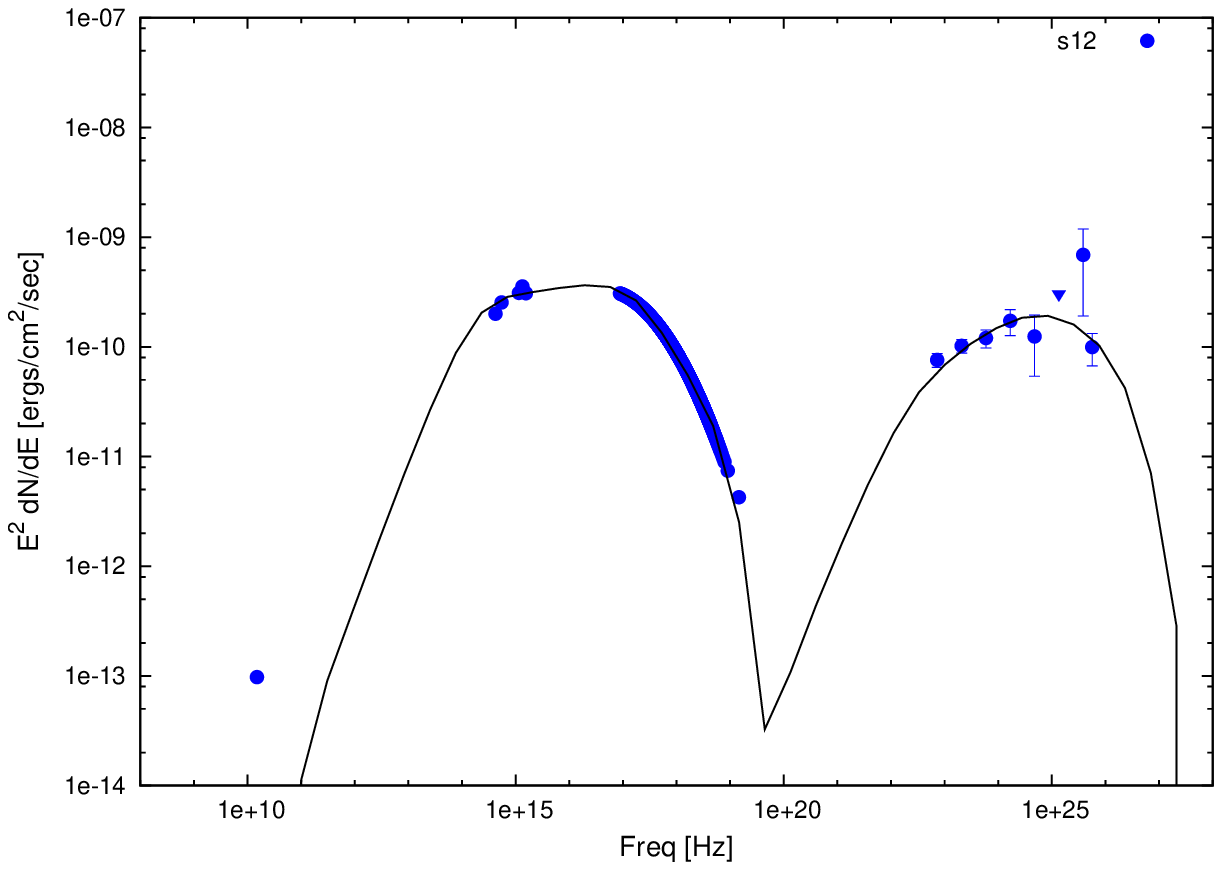}}
      \qquad
  	\subfloat[s13]{\includegraphics[scale=0.6]{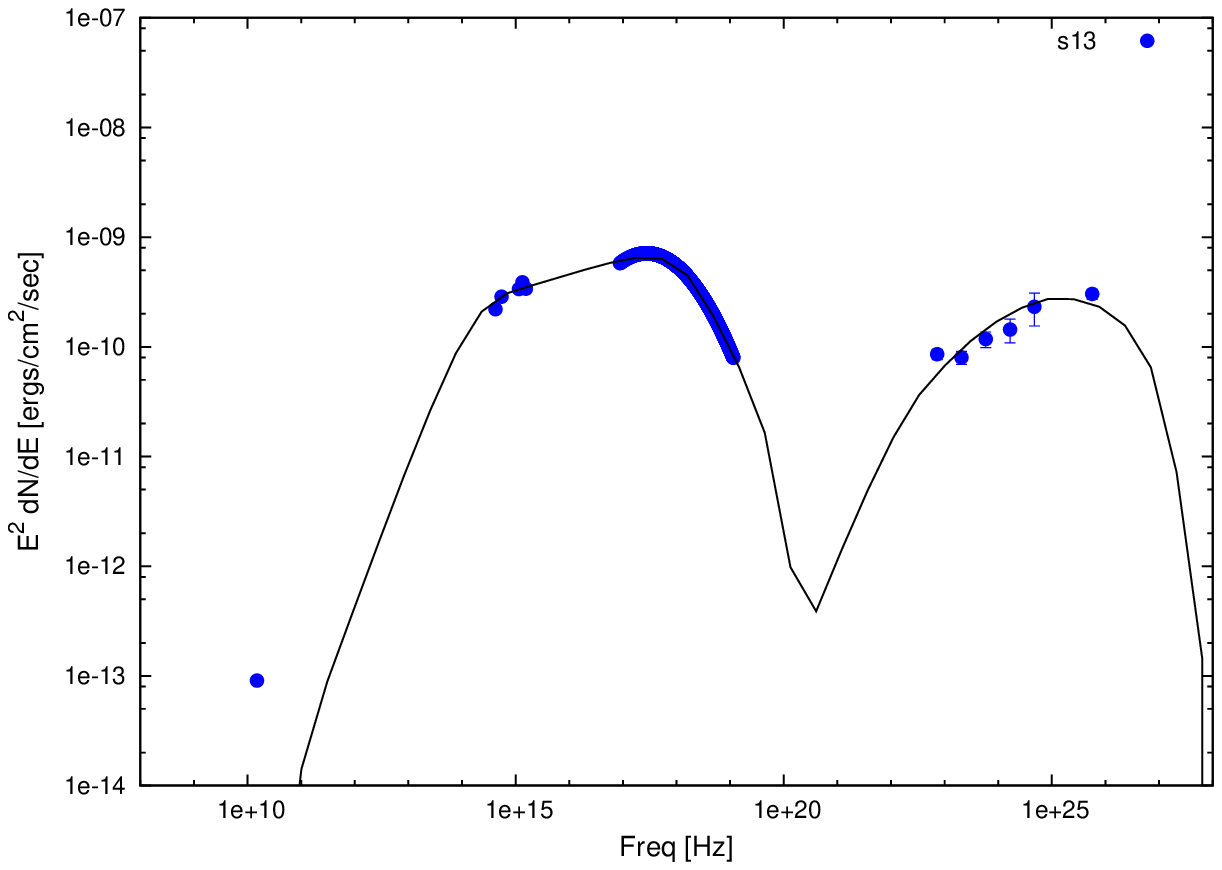}}
      \qquad
  	\subfloat[s14]{\includegraphics[scale=0.6]{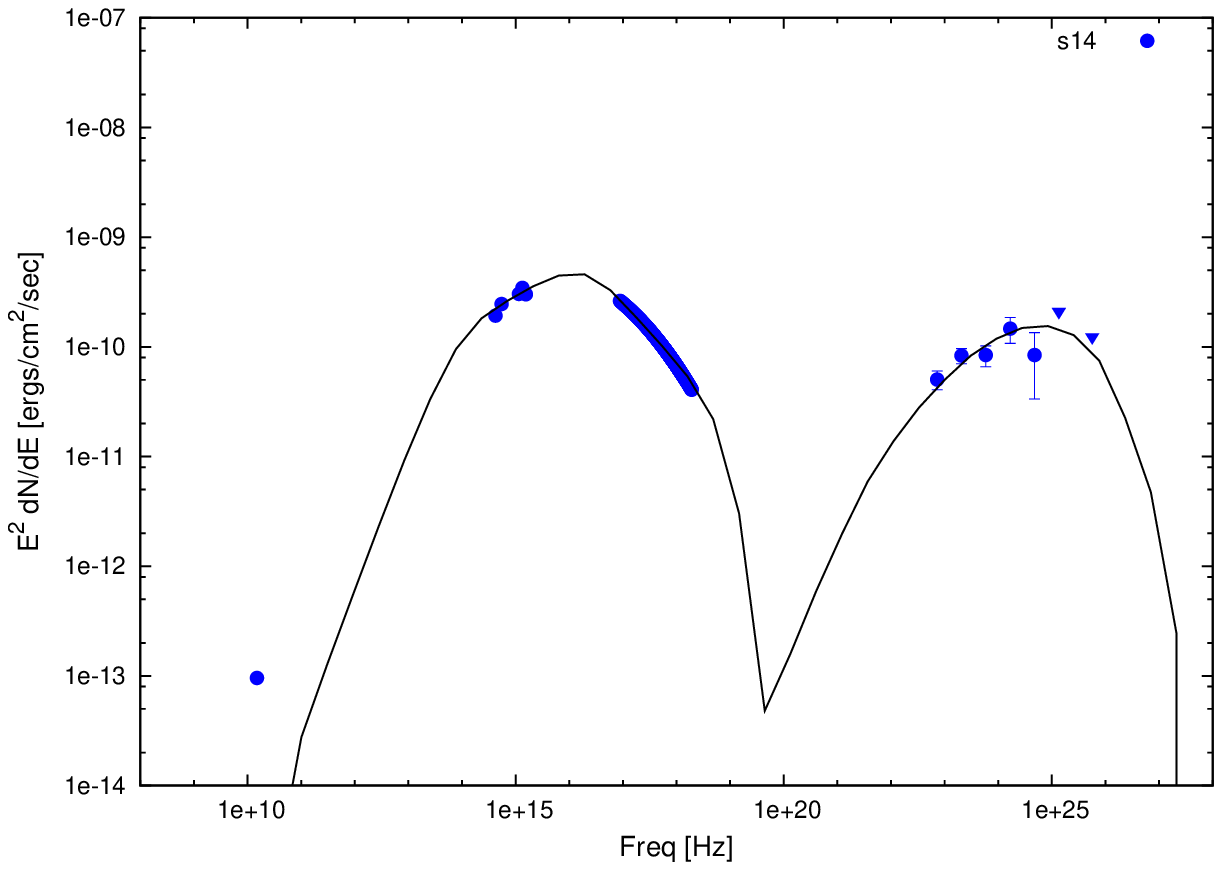}}
      \qquad
  	\subfloat[s15]{\includegraphics[scale=0.6]{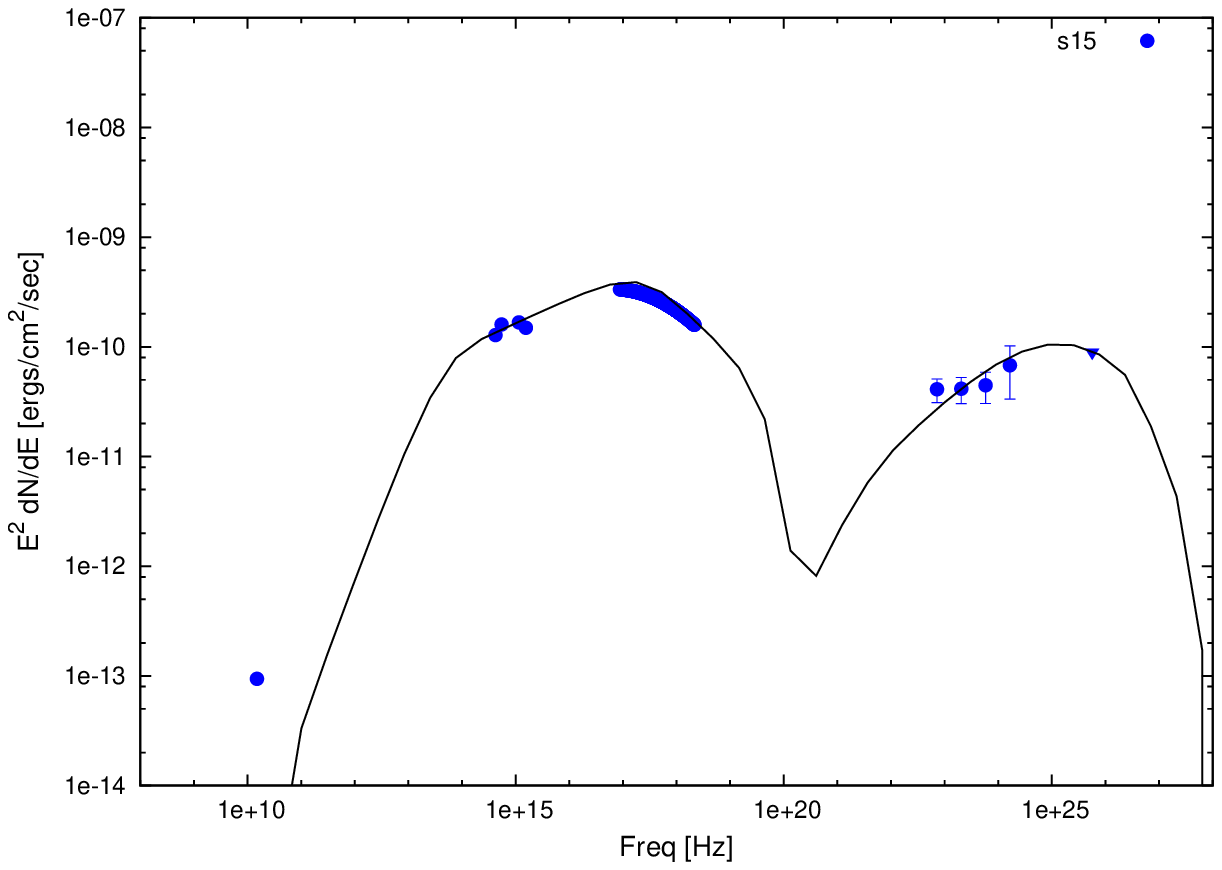}}
      \qquad
  	\subfloat[s16]{\includegraphics[scale=0.6]{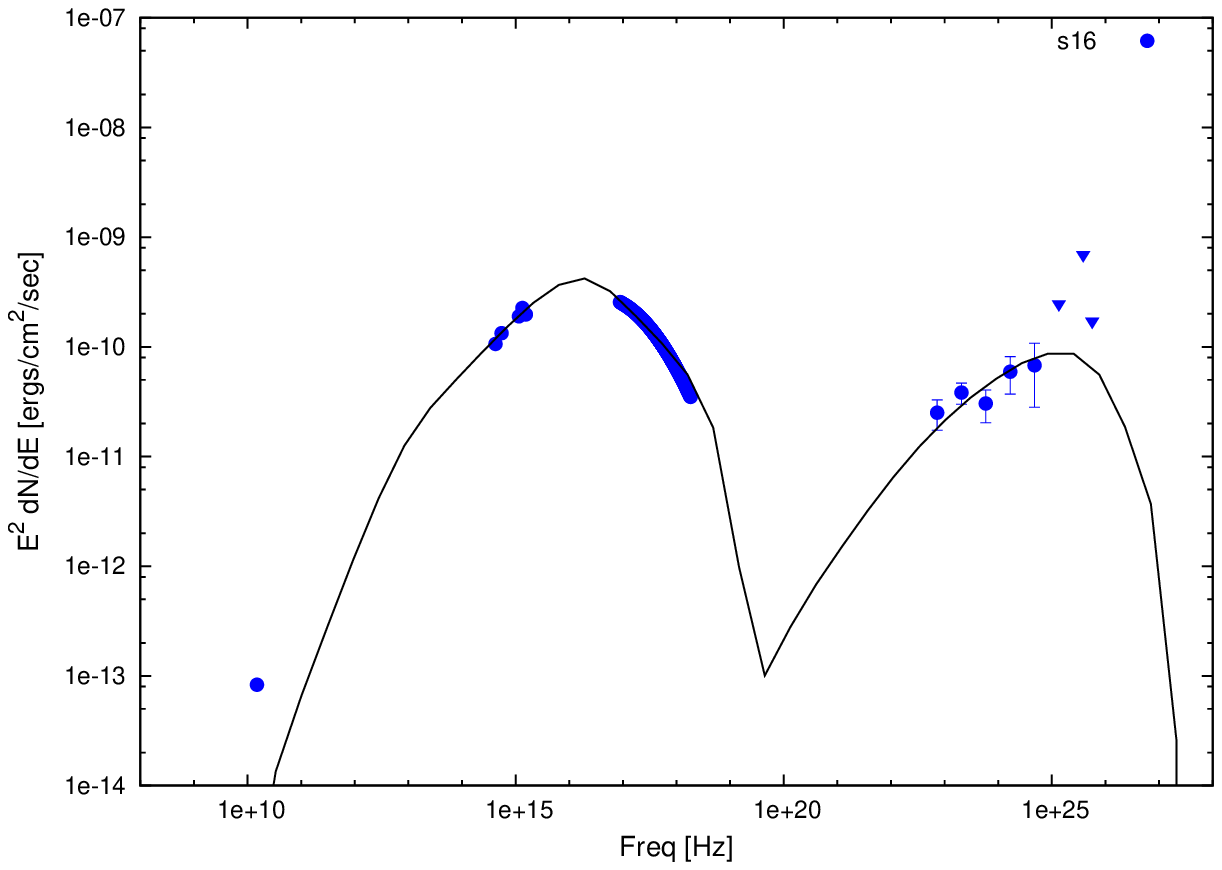}}
      \qquad
  \end{figure*}

\end{document}